\documentclass[aps,pra,reprint,superscriptaddress,showpacs]{revtex4-1}
\usepackage[utf8]{inputenc}
\usepackage{color}
\usepackage{graphicx}
\usepackage{amssymb,amsmath,amsthm,amsfonts}
\usepackage{mathtools,physics}
\usepackage{float}
\usepackage{hyperref}
\usepackage{enumitem}
\usepackage{filecontents}

\usepackage{url}

\begin{document}

\title{Pulse optimization for high-precision motional-mode characterization in trapped-ion quantum computers}

\author{Qiyao Liang}
\email{qiyao@mit.edu}
\affiliation{Duke Quantum Center, Duke University, Durham, NC 27701, USA}
\affiliation{Department of Electrical Engineering and Computer Science, Massachusetts Institute of Technology, Cambridge, MA 02139, USA}
\author{Mingyu Kang}
\affiliation{Duke Quantum Center, Duke University, Durham, NC 27701, USA}
\affiliation{Department of Physics, Duke University, Durham, NC 27708, USA}
\author{Ming Li}
\email{Current affiliation: Atom Computing, Inc., USA}
\affiliation{19 Pomona Ave, El Cerrito, CA 94530, USA}
\author{Yunseong Nam}
\affiliation{Department of Physics, University of Maryland, College Park, MD 20742, USA}
\date{\today}

\begin{abstract}
  High-fidelity operation of quantum computers requires precise knowledge of the physical system through characterization. For motion-mediated entanglement generation in trapped ions, it is crucial to have precise knowledge of the motional-mode parameters such as the mode frequencies and the Lamb-Dicke parameters. Unfortunately, the state-of-the-art mode-characterization schemes do not easily render the mode parameters in a sufficiently accurate and efficient fashion for large-scale devices, due to the unwanted excitation of adjacent modes in the frequency space when targeting a single mode, an effect known as the \textit{cross-mode coupling}. Here, we develop an alternative scheme that leverages the degrees of freedom in pulse design for the characterization experiment such that the effects of the cross-mode coupling is actively silenced. Further, we devise stabilization methods to accurately characterize the Lamb-Dicke parameters even when the mode frequencies are not precisely known due to experimental drifts or characterization inaccuracies. We extensively benchmark our scheme in simulations of a three-ion chain and discuss the parameter regimes in which the shaped pulses significantly outperform the traditional square pulses.
\end{abstract}

\maketitle

\section{Introduction}
\label{sec:intro}
With the advent of promises to deliver a quantum computer of scale beyond the capability of conventional computers, comes the responsibility to indeed manufacture one that can reliably be tuned for its normal operations. 
Without the ability to tune, a quantum computer may produce unpredictable and erroneous results. This necessitates the development of methods that can efficiently and accurately characterize the system parameters, as a prerequisite for its high-fidelity operations.

In this paper, we aim to address this challenge for a trapped-ion quantum computer. In particular, our focus is to enhance the accuracy of motional-mode parameter estimation~\cite{Mingyu2022}, as these parameters are essential for designing and performing high-fidelity two-qubit operations~\cite{PhysRevLett.74.4091, PhysRevLett.82.1835, PhysRevLett.82.1971, Blumel21, blumel2021power, li2022realizing}. An accurate characterization of mode parameters can provide several additional benefits as well. First, the overhead in gate calibrations caused by incorrect parameter estimates~\cite{maksymov2021optimal, Gerster21} can be reduced. Second, gate pulses can be more efficient in terms of control-signal power, gate duration, and robustness than pulses designed with inaccurate motional-mode parameters~\cite{blumel2021power, Blumel21}.

In a characterization experiment, a model is used to predict the physical parameters being characterized via fitting to some measured quantities. A major bottleneck to achieving high-accuracy characterization is the mismatch between the model's predictions and experimental outcomes~\cite{Mingyu2022}, resulting from various sources of experimental imperfections and/or model violations.  Ref.~\cite{Mingyu2022} presents improved fitting models as well as parallelized characterization protocols to address some of these mismatches. To achieve further improvement in the accuracy of mode-parameter characterization, here, we investigate a complementary approach to the characterization tools developed in~\cite{Mingyu2022} by leveraging the control degrees of freedom that are already readily available for the quantum gate implementation on the hardware level.

Specifically, we use pulse optimization, which is a widely used tool in quantum optimal control~\cite{qoct}. Indeed, many pulse-shaping paradigms have been developed, in particular for the purpose of designing high-fidelity single-qubit and two-qubit gates~\cite{grape, crab, krotov}. For trapped ions, the pulse shaping techniques are standard and have been previously adopted to accomplish amplitude-modulated (AM)~\cite{Zhu06, Roos08, Kim09, Choi14, Debnath16, Figgatt19}, frequency-modulated (FM)~\cite{James, Landsman19, Wang20, brobust, frobust}, and phase-modulated (PM)~\cite{Shapira18, Milne20, Green15, Lu19} entangling gates.
Here, we devise pulse shaping paradigms to eliminate the effect of dominant error sources in the characterization process, which in turn can help improve the performance of multi-qubit gates~\cite{katz2023programmable, katz2023demonstration} and parallel gates~\cite{Figgatt19, Bentley20, Grzesiak20}. We note that a recent work explored pulse shaping for system parameter characterization in a single-qubit setup~\cite{OptimizedBayes}.

The rest of the paper is structured as follows.
In Sec.~\ref{sec:prelim}, we provide necessary preliminary information for concretely defining our mode characterization problem.
We then formalize the problem in
Sec.~\ref{sec:problem}, where we focus on the cross-mode coupling and detuning
errors, to be detailed in the section, as the two dominant 
sources of error in the mode characterization.
In Sec.~\ref{sec:method}, we explore pulse shaping, 
where we exploit the degrees of freedom in modulating 
the beams that illuminate the ions to engineer
the response of the system, such that the previously discussed two errors are minimized. In Sec.~\ref{sec:results} we demonstrate the viability of our proposed pulse shaping schemes via extensive numerical simulations on a three-ion chain.
Finally, we discuss the applications and potential limitations of our proposed scheme
in Sec.~\ref{sec:discussion} and provide outlook in Sec.~\ref{sec:conclusion}.

\section{Preliminaries}
\label{sec:prelim}
In a typical trapped-ion quantum computer, a qubit is encoded in two internal states of an ion. As $N$ ions form a linear Coulomb crystal, the collective motion of the ions can be decomposed into $3N$ normal modes, each of which is described as a quantum harmonic oscillator. A pair of qubits within the ion chain can be entangled via a state-dependent force, frequently using lasers. Such force couples the ions' internal states to the motional modes, typically strongly coupling to a subset of $N'$ modes, where the motional modes serve as a ``bus'' that establishes the communication between the target pair of ions that we want to entangle. 

To produce an ideal entangling gate, two criteria must be satisfied; first, the residual entanglement between the qubits and the motional modes must be removed at the end of the gate; second, the degree of entanglement between the target pair of qubit states must reach a pre-specified amount. 
Indeed, accurate knowledge of the motional-mode parameters, mode frequencies and mode-ion coupling strengths (Lamb-Dicke parameters), is necessary to exactly satisfy the criteria. 
In practice though, the parameters are inevitably known only to finite accuracy, due, e.g., to faulty characterization or experimental drifts. This necessitates gate pulse design methods to construct solutions that are robust to parameter offsets, albeit at the cost of a significantly higher power or longer pulse length~\cite{blumel2021power, Blumel21, frobust, Shapira18, Milne20, arobust, Valahu22}. Despite much effort, readily available rough theoretical estimates of the mode parameters, calculated using minimal knowledge of the system configurations such as the voltages applied to the trap~\cite{NIST_Bible}, have been proven to be insufficient for high-fidelity quantum operations, as evidenced in Ref.~\cite{Blumel21} (see also footnote [35] of Ref.~\cite{Mingyu2022}). 

It is thus not surprising that spectroscopic methods~\cite{Mavadia14, Goodwin16, Stutter18, Hrmo18, Welzel18, Joshi19, Hrmo19, Jarlaud20, Feng20, chen2020efficient, Sosnova21} have been employed to better estimate the mode parameters. Briefly, in all these methods, one starts by initializing the ion’s internal and motional state to the ground state $|0,0\rangle_{j,p^*}$ following the standard initialization and cooling procedures. Here $|a,b\rangle_{j,p^*}$ denotes the composite state of the computational basis state $|a\rangle_j$ of ion $j$ with $a\in\{0,1\}$ and the motional Fock state $|b\rangle_{p^*}$ of mode $p^*$ with phonon number $b$, and $p^*$ is the index of the mode we aim to characterize. One then turns on the Hamiltonian
\begin{widetext}
\begin{align}
\hat{H}_{I,j}(t) &= \hat{\sigma}^+_j \exp{i\sum_{p=1}^{N'} \eta_{j,p} (\hat{a}_p e^{-i\omega_p t} + \hat{a}_p^\dagger e^{i\omega_p t})} g_j(t) + h.c.,
\label{eq:HI}
\end{align}
\end{widetext}
by illuminating ion $j$ for a duration $\tau$ with the pulse form $g_j(t)$ that in principle can be of any shape, where $\hat{\sigma}^+_j := \left|1\right> \left<0\right|$ acts on the computational basis state of ion $j$, $\hat{a}_p^\dagger$ and $\hat{a}_p$ are the creation and annihilation operators acting on mode $p$, $\omega_p$ is the frequency of mode $p$, and $\eta_{j,p}$ is the Lamb-Dicke parameter that couples ion $j$ and mode $p$, defined according to
$
    \eta_{j,p} := b_{j,p} |\vec{k}_{j,p}|/\sqrt{2m \omega_p},
$
where $b_{j,p}$ is the $j$-th element of the normalized eigenvector of mode $p$, $m$ is the ion mass, and $\vec{k}_{j,p}$ is the wavevector of the electric field that couples ion $j$ to mode $p$, projected along the motional direction of mode $p$. Here we assume $|g(t)|$ to be much smaller than the qubit frequency splitting.

Indeed in~\cite{Mavadia14, Goodwin16, Stutter18, Hrmo18, Welzel18, Joshi19, Hrmo19, Jarlaud20, Feng20, chen2020efficient, Sosnova21} $g_j(t)$ was chosen to be near-resonant to one of the blue-sideband (BSB) transition frequencies $\omega_{p^*}$, i.e., $g_j(t)={A}_j\exp\{-i (\mu_j t + \phi_j)\}$ with the detuned drive frequency $\mu_j = \tilde{\omega}_j - \omega_j^{\rm{qbt}}$ $\approx$ $\omega_{p^*}$, where $A_j$ is the Rabi frequency of the resonant qubit-state transition, $\phi_j$ is the laser phase, $\tilde{\omega}_j$ is the frequency of the laser's electric field, and $\omega_j^{\rm{qbt}}$ is the frequency separation between the two internal states of qubit $j$.
The ion-mode system then undergoes the so-called BSB transition, where a Rabi flop between the states $|0,0\rangle_{j,p^*}$ and $|1,1\rangle_{j,p^*}$ occurs. Since the final $|1\rangle_j$-state population of ion $j$ at time $\tau$ has dependence on $\omega_{p^*}$ and $\eta_{j, p^*}$ according to (\ref{eq:HI}), one can then extract the values of the mode parameters by measuring the population through statistics accumulated over multiple shots and comparing the theoretical predictions implied by (\ref{eq:HI}) with the measurements.  

We consider the aforementioned, conventional spectroscopy method using $g_j(t)={A}_j\exp\{-i \mu_j t + i \phi_j\}$ as our baseline, with which we compare our characterization method using pulse shaping, to be developed in later sections. Note the baseline uses a single-tone pulse, where the drive frequency $\mu_j$ is near-resonant with the frequency $\omega_{p^*}$ of the target mode $p^*$. We denote such non-modulated pulses as \textit{square pulses}.

\section{Challenges to Accurate and Efficient Mode Characterization}
\label{sec:problem}
In theory, if one can simulate (\ref{eq:HI}) efficiently and exactly, the mode parameters $\omega_p$ and $\eta_{j,p}$ for all $j$ and $p$ can in principle be estimated to within the model violation. However, such a simulation becomes a computationally prohibitive task as the system size increases. Traditionally, the so-called single-mode BSB model~\cite{NIST_Bible} has thus been employed to enable a scalable simulation, effectively focusing on a single mode and a single ion only, at the cost of introducing further, non-insignificant model-violation errors. 

In practice, additional considerations need to be given. Recall the agreement between the model predictions and the experimental results in the BSB transition population is at the core of the mode-parameter estimation and the population has dependencies on both $\omega_p$ and $\eta_{j,p}$. Inaccurate estimation of $\omega_p$, a known parameter to fluctuate and drift~\cite{frobust}, would thus result in inherent inaccuracy in estimating $\eta_{j,p}$ and vice versa.

To resolve the theoretical and practical issues then, two complementary approaches may be considered. The first approach involves finding more sophisticated theoretical models that better approximate the system while still remaining computationally feasible and developing tailor-made experimental protocols that better separate the effect of the uncertainties in $\omega_p$ from that of the uncertainties in $\eta_{j,p}$ in the various BSB populations induced empirically. This approach has indeed been studied in detail in Ref.~\cite{Mingyu2022}, however using a square pulse. In a contrasting and complementary approach, which is the one explored and investigated here, active ``noise canceling'' can be considered, via \textit{pulse shaping}: By deliberately constructing a non-square, \textit{shaped pulse} to eliminate significant model-violation terms or desensitize the BSB-transition population with respect to parameter change, significant improvement in the mode-parameter estimation can be achieved. 

In this section, we concretely lay out the necessary technical details for understanding our pulse shaping method and its benefits. By showing the approximations used to arrive at the simple, single-mode BSB model, we can pinpoint the terms responsible for the model-violation errors. Further, we can explicitly express the sensitivity term. Note, in the forthcoming discussion, for simplicity, we consider the case of illuminating one ion at a time for one target mode, thus omitting the reference to ion $j$ wherever contextually clear; Parallelization by simultaneously illuminating multiple ions and targeting multiple modes~\cite{Mingyu2022} may be considered but is beyond the scope of this paper. 

To start, we reiterate that, when the mode characterization is performed for a trapped-ion quantum computer, experimental evolution is compared to that implied by the simulated Hamiltonian. Specifically, a series of experimentally observed BSB-transition population $P$ is compared with that expected from the simulation, denoted ${\mathcal P}$, to reveal the mode parameters of interest. As a result, the discrepancy between $P$ and $\mathcal{P}$ directly leads to errors in characterizing the mode parameters. 

{\it Multi-mode and Single-mode BSB Hamiltonian} -- To obtain $\mathcal{P}$ at scale beyond a handful number of ions, we first apply a series of approximations to the Hamiltonian in (\ref{eq:HI}) to arrive at the single-mode BSB Hamiltonian, better amenable to a scalable simulation. Assuming small $\eta_p ({\leftarrow} \eta_{j,p})$, we linearize the Hamiltonian in (\ref{eq:HI}) by performing a first-order Taylor expansion on the exponential term in (\ref{eq:HI}). Then, a series of rotating-wave approximation (RWA) is applied to remove the frequency components that are far off-resonant from all mode frequencies (see Appendix~\ref{app:derivations} for details). Here we assume a more general form of the pulse function $g(t)$ that have frequency componenet(s) near the mode frequencies only. These approximations combine to give the $N'$-mode BSB Hamiltonian
\begin{equation} \label{eq:HNmode}
\hat{H}'_I(t) = i \sum_{p=0}^{N'-1} \eta_{p} e^{i \omega_p t} g(t) \hat{\sigma}^+ \hat{a}^\dagger_p + h.c.
\end{equation}
Next, as only mode $p^*$ is targeted, we may apply a stronger RWA to further remove the frequencies of the non-target modes, assuming $|\eta_p A|\ll |\omega_p-\mu|\;\forall p\neq p^*$. This results in the single-mode BSB Hamiltonian
\begin{equation} \label{eq:H1mode}
\hat{H}'_{I,p^*}(t) = i \eta_{p^*} e^{i \omega_{p^*} t} g(t) \hat{\sigma}^+ \hat{a}^\dagger_{p^*} + h.c.
\end{equation}
See detailed derivations and approximations used to arrive at \eqref{eq:HNmode} and \eqref{eq:H1mode} in Appendix~\ref{app:derivations}. We note that the simpler model in (\ref{eq:H1mode}) demands far less computational cost in its simulation, whereas the cost for simulating the more complicated model in (\ref{eq:HNmode}) grows exponentially with the number of modes $N'$. Nonetheless, the second RWA that we apply leading to \eqref{eq:H1mode} no longer holds when $|\eta_p A|$'s are of comparable magnitudes to the finite mode-frequency spacings $|\omega_{p^*}-\omega_p|$, which constitutes the heart of the problem that we describe next.

We are now ready to discuss our problem details that base the comparison between $P$ and ${\mathcal P}$, both of which we obtain by simulation in this paper. Starting with $P$, recall the most complete theoretical model considered herein is (\ref{eq:HI}). Comparing (\ref{eq:HI}) and (\ref{eq:HNmode}), one can see that the leading-order difference is $O(\eta^2)$, which is indeed the well-known Debye-Waller (DW) effect~\cite{Wineland79, NIST_Bible}. Briefly, it affects the population $P$ by reducing the effective BSB Rabi frequency -- the spread of the ion's position wavepacket, captured in (\ref{eq:HI}) but not in (\ref{eq:HNmode}), leads to this reduction. Since our goal is to silence the leading-order, significant model-violation terms, which we show to be $O(\eta)$ in the paragraphs to come later, for our purposes, it suffices to consider the evolution implied by (\ref{eq:HNmode}) as our simulated experiment that generates the (simulated) experimental population $P$. 
Further, the DW effect is known to be re-capturable by using a more advanced model that predicts the BSB population without adding significant computational cost~\cite{Mingyu2022}.
As for the model-based, approximate BSB population ${\mathcal P}$, we obtain it from simulating the evolution implied by the Hamiltonian in (\ref{eq:H1mode}) over duration $\tau$.

{\it Elimination targets for pulse shaping} -- The source of discrepancy between $P$ and ${\mathcal P}$
forms the elimination target by our pulse shaping. Such an elimination is in general not achievable via a simple square pulse. By bringing $P$ and ${\mathcal P}$ closer via shaping the pulses, we can enable significantly better mode-parameter estimation. 
To this end, the elimination targets are: 

\textit{(1) Cross Mode Coupling (CMC) error}: Note an important difference between (\ref{eq:HNmode}) and (\ref{eq:H1mode}) is that former captures the coupling of internal ion states to modes $p\neq p^*$, an effect hereafter referred to as the CMC, whereas the latter does not. Specifically, the leading-order CMC error may be quantified as
\begin{equation}
    \theta^{\rm CMC}_{p} = \Theta_{p}^{(1)}
    := \int_0^{\tau} g(t) e^{i\omega_p t} dt \quad\quad (p \neq p^*),
    \label{eq:cmc}
\end{equation}
where $\Theta_{p}^{(1)}$ is the first-order Magnus integral,
since the first-order Magnus-approximated evolution operator $\hat{U}^{(1)}$ for (\ref{eq:HNmode}) is
\begin{equation} \label{eq:uni_approx}
\hat{U}^{(1)} := \exp (\hat{\Omega}_1)=
\exp \left( \sum_{p=0}^{N'-1} \eta_p \Theta_{p}^{(1)} \hat{\sigma}^+ \hat{a}_p^\dagger - h.c. \right)
\end{equation}
and we used $\hat{\Omega}_1$ to denote the first-order Magnus term.
Note this approximation is valid when $|\eta_{p^*} g(t)|$ is small compared to $1/\tau$.
By eliminating all $\theta_p^{\rm CMC}$ defined in (\ref{eq:cmc}) 
using pulse shaping, the first-order CMC effect may be suppressed, leaving the second-order Magnus term
\begin{equation}
    \hat{\Omega}_2 = \sum_{p,p'=0}^{N'-1} \eta_p \eta_{p'} \Theta_{p,p'}^{(2)} 
\left(\hat{\sigma}_z \hat{a}^\dagger_p \hat{a}_{p'} + \hat{\sigma}_- \hat{\sigma}_+ \delta_{p,p'} \right)
- h.c.,
\end{equation}
where
\begin{equation}
    \Theta_{p,p'}^{(2)} = \frac{1}{2} \int_0^\tau dt_1 \int_0^{t_1} dt_2 
    g(t_1) g^*(t_2) \: e^{i\omega_p t_1} \: e^{-i \omega_{p'} t_2}
\label{eq:2ndMag}
\end{equation}
is the second-order Magnus integral, as the post-suppression leading-order CMC error.

\textit{(2) Detuning error}: In our model Hamiltonian, $\omega_p$ are considered static. However, experimental imperfections, either from faulty characterization or drifts during the experiment~\cite{Maksymov2022}, give rise to inexactness or detuning of the mode frequencies $\omega_p$. This manifests in our model as
\begin{align}
    \theta^{\rm{det}}_p (\delta_p) &:= \int_0^{\tau} e^{i(\omega_p + \delta_p)t}g(t)dt - \int_0^{\tau} e^{i\omega_{p}t} g(t)dt \nonumber\\
    &= \sum_{\kappa=1}^\infty \frac{\delta_p^\kappa}{\kappa!} \frac{\partial^\kappa}{\partial \omega_p^\kappa} \Theta_{p}^{(1)},
    \label{eq:det}
\end{align}   
which is the difference incurred in the first-order Magnus integral due to the detuning $\omega_p \rightarrow \omega_p + \delta_p$. 
For the target mode $p^*$, the difference leads to off resonance of the BSB transition, which directly causes error in estimating $\eta_{j,p^*}$. For the non-target modes, the detuning leads to changes in the CMC terms in (\ref{eq:cmc}), hence imperfect removal of the CMC effects via pulse shaping, which ultimately results in the characterization error for $\eta_{j,p^*}$. Since the precise value of $\theta_p^{\rm{det}}$ cannot be learned when $\delta_p$ is not known precisely, we aim to suppress $\theta_p^{\rm{det}}(\delta_p)$ via minimizing the first $K$ leading-order derivatives ($\kappa \leq K$) of $\Theta_p^{(1)}$ within the Taylor expansion terms of (\ref{eq:det}).

To recap, we emphasize that the quantities (\ref{eq:cmc}) and (\ref{eq:det}) are in general nonzero for the non-modulated pulse of the form $g(t) = \bar{A}e^{-i \mu t + i\phi}$. In our work, we deliberately design $g(t)$ to remove (\ref{eq:cmc}) and (\ref{eq:det}), such that the accuracy of mode characterization is significantly improved. Note that the residual errors in the characterization would then be due to the contributions from the second- and higher-order Magnus terms as mentioned earlier, in addition to other high-order Hamiltonian terms unaccounted for in our model. 
Indeed, an example may be the aforementioned DW effect. In the rest of our paper, we therefore neglect the DW effect and focus on the larger, leading-order effects such as the CMC and detuning error that we actively silence using our pulse shaping.

\section{Method}
\label{sec:method}

In this section, we present pulse-shaping techniques for suppressing the CMC and the detuning errors by removing (\ref{eq:cmc}) and (\ref{eq:det}), respectively. To summarize, we aim to obtain the following goals:
\begin{enumerate}[label=Goal \arabic*]
\item Suppress the CMC by imposing $\Theta_{p}^{(1)}=0$ for all $p \neq p^*$; 
\label{goal1}
\item Maximize $|\Theta_{p^*}^{(1)}|$ while keeping the average Rabi frequency of the pulse the same, such that maximal change in the qubit population (observable in our experiment) is reached with a reasonable power requirement;
\label{goal2}
\item Achieve \ref{goal1} and \ref{goal2} and estimate $\eta_{j,p^{*}}$ well even in the presence of uncertainties in $\omega_p$'s by stabilizing the conditions with respect to changes in $\omega_p$'s.
\label{goal3}
\end{enumerate}

Briefly, to eliminate the CMC error, we first assume $\omega_p$'s are fully known, a condition which we relax later when introducing the detuning errors. Our \ref{goal1} is then to obtain $\theta_p^{\rm CMC} = 0$ in (\ref{eq:cmc}) for all $p \neq p^*$. This way, the non-target modes are decoupled from the illuminated ion up to the first order. 

In principle, a pulse that achieves \ref{goal1} can, for example, be far detuned from all of the mode frequencies. This would result in a minimal target-mode signal (qubit population) or, said differently, a pulse solution with a prohibitively large power. To address this, we maximize the target mode's response (\ref{goal2}) by maximizing $|\Theta_{p^*}^{(1)}|$ while keeping the pulse power requirement constant, to be more concretely defined later.

Relaxing now the perfect knowledge assumption for $\omega_p$'s, we notice that the precise integral values $\Theta_{p}^{(1)}$ used to achieve \ref{goal1} and \ref{goal2}, for both $p{=}p^*$ and $p{\neq}p^*$, will change if a detuning error occurs. To prevent a significant change, we can stabilize all $\Theta_{p}^{(1)}$'s with respect to their respective mode frequencies $\omega_p$'s such that $\theta_p^{\rm det}$ in (\ref{eq:det}) is suppressed to (near) zero for a small range of detuning, with a varying degree of stabilization $K$. In other words, for \ref{goal3}, we can minimize the derivatives of $\Theta_{p}^{(1)}$ with respect to $\omega_p$ for all modes $p$. 

In what follows, we focus on achieving all three goals stated above by appropriately shaping $g(t)$.
To start, we expand $g(t)$ using a Fourier basis, i.e.,
\begin{equation}
\label{eq:Fourier}
g(t) = \sum_n A'_n e^{-i \frac{2\pi n}{\tau}t+i\phi_n} = \sum_n A_n e^{-i \frac{2\pi n}{\tau}t},
\end{equation}
where $n$ indexes the Fourier basis and $A_n := A_n'e^{i\phi_n}$ is a complex coefficient of each Fourier component. While a complete Fourier basis comprising an infinite number of basis components may be considered, in practice, a Fourier basis consisting of $N_{\rm{basis}}$ components, concentrated around motional mode frequencies $\omega_p$'s, suffices. 
The average Rabi frequency $\bar{A}$ of pulse $g(t)$ is then defined as $\bar{A} := \sqrt{\sum_n |A_n|^2}$.
Inserting (\ref{eq:Fourier}) into $\Theta_{p}^{(1)}$ defined in (\ref{eq:cmc}), we obtain 
\begin{align}
   \Theta^{(1)}_p = \sum_n A_n\int_0^\tau \exp \left\{ i 
   \left(\omega_p - \frac{2\pi n}{\tau}
   \right)t \right\}dt ,
   \label{eq:TP1}
\end{align}
which we can then summarize into a $N'\times 1$ column vector $\vec{\Theta}^{(1)}$. Extracting the Rabi frequency coefficients $A_n$ from the above expression and summarizing again as a column vector $\vec{A}$, we can succinctly express $\vec{\Theta}^{(1)}$ as
\begin{align}
\vec{\Theta}^{(1)} = \mathbf{M} \vec{A},
\label{eq:firstordermagnus}
\end{align}
where $\mathbf{M}$ is a matrix of dimension $N'\times N_{\rm{basis}}$ with the matrix elements
\begin{align}
M_{p,n} = \int_0^\tau  \exp \left\{ i \left(\omega_p - \frac{2\pi n}{\tau}\right)t  \right\} dt. 
\end{align}
The goal here becomes then to determine the appropriate Fourier expansion coefficients $\vec{A}$ that achieve the three goals listed above. 

\textit{Step 1 (\ref{goal1}):} {\it CMC suppression --} Here, we aim to ensure the inner product between each row of $\mathbf{M}$ and $\vec{A}$ is zero for all $p{\neq}p^*$, such that $\theta_{p}^{\text{CMC}}{=}0, \, \, \forall p{\neq}p^*$.
To achieve this, we remove the $p^*$-th row vector from $\mathbf{M}$, denote this matrix as $\mathbf{M}'$, and find the null space of $\mathbf{M}'$. We characterize this space  $\boldsymbol{\mathcal{S}}^{(1)}$ by a spanning set of vectors $\mathcal{S}^{(1)}$, and denote $\mathbf{N}$ as the matrix whose columns consist of vectors in $\mathcal{S}^{(1)}$. These nullspace vectors then satisfy the CMC nulling conditions of all $p\neq p^*$ and serve as the spanning vectors of the space over which we maximize the signal of the target mode $p=p^*$. The size of the nullspace is the number of basis functions, subtracted the number of nulling conditions applied $N'-1$. The null-space dimension marks the degrees of freedom we have left to work with for the signal maximization, of which there needs to be a sufficient number to warrant convergence in the pulse shape. This implies that the pulse length has to be sufficiently long such that the frequency resolution $2\pi n/\tau$ is small enough to generate sufficiently many frequency basis components.

\textit{Step 2 (\ref{goal2}):} {\it Maximizing signal strength --} To maximize the signal strength of the target mode $p^*$, we first minimize the average power of the pulse to achieve $|\Theta_{p^*}^{(1)}|=1$. 
We then scale the obtained, minimal-power pulse solution vector by \textit{Rabi-frequency scaling factor} $\alpha$, such that $|\Theta_{p^*}^{(1)}|$ reaches a desired value, in this case $\alpha$. 
More specifically, our starting point is the projection of the $p^*$-th row vector $\vec{v}^\dagger$ of $\mathbf{M}$ onto the nullspace $\mathbf{N}$. We then find the eigenvector $\vec{\xi}$ with the largest eigenvalue $\lambda_{\rm max}>0$ of the positive-definite matrix $\mathbf{N}^\dagger \vec{v} \vec{v}^\dagger \mathbf{N}$. The normalized pulse solution is obtained by projecting $\vec{\xi}$ onto the nullspace $\mathbf{N}$ then normalizing by $\sqrt{\lambda_{\rm max}}$.
After rescaling by a factor of $\alpha$, we obtain the pulse solution $\vec{A}$, given by
\begin{align}\label{eq:normalizepulse}
    \vec{A} = \frac{\alpha}{\sqrt{\lambda_{\rm max}}} \mathbf{N}\vec{\xi}.
\end{align}
It is worth mentioning that our choice of $\alpha$ is restricted by the conditions of the RWA approximations. As a result, we mainly operate in the small $\alpha$ regime.

\textit{(Optional) Modified Step 1 (\ref{goal3}):} {\it Stabilization against detuning -- } We now admit that we have inaccuracies in $\omega_p$'s such that $\omega_p \rightarrow \omega_p + \delta_p$, $\forall p$. The resulting error contribution $\theta_p^{\rm det}$, given in (\ref{eq:det}), can be suppressed by adding the following constraints to the null-space conditions:
\begin{equation}
\frac{\partial^\kappa}{\partial \omega_p^\kappa}\Theta_{p}^{(1)}= 0
\quad\quad \forall p, \: \kappa \in \{1,..,K\},
\label{eq:stab}
\end{equation}
where $K$ is the \textit{moment of stabilization}, which corresponds to the highest order of derivatives that are nulled.

Specifically, these constraints are appended as additional $KN'$ rows to the matrix $\mathbf{M}'$, where each row contains the nulling condition for each pair of $p$ and $\kappa$. As $\mathbf{M}'$ originally has $N'-1$ rows, the appended matrix entries in the expanded $\mathbf{M}'$ are given by
\begin{align}
    M'_{\kappa N' - 1 + p, n}=\frac{\partial^\kappa}{\partial\omega_p^\kappa}\int_0^\tau  \exp \left\{ i (\omega_p - 2\pi n/\tau) t \right\} dt,
\end{align}
where $p = 1,..,N'$ and $\kappa = 1,..,K$. Then, we find the set of null space vectors $\mathcal{S}^{(1)}$ corresponding to the expanded $\mathbf{M}'$ and proceed to perform the same signal maximization procedure as described in \textit{Step 2}. 

We note in passing that performing stabilization against detuning as described in \textit{Modified Step 1} is optional. Should the rough $\omega_{p}$ values known a priori to the characterization be sufficiently accurate for the purpose of characterization, the need for these constraints are obviated. 

{\it Remarks} -- Due to its role in the rescaling of $|\Theta_{p^*}^{(1)}|$, hereafter, we use $\alpha$ interchangeably as the absolute value of the first-order Magnus integral for mode $p^*$ for both shaped and square pulses. 
The rescaled pulse roughly induces a qubit population inversion $\mathcal{P} \approx \sin^2 (\eta_{p^*} |\Theta_{p^*}^{(1)}|)$ (see Appendix \ref{sec:appendix_population} for details) and, in the perturbative regime where $|\eta_{p^*} \Theta_{p^*}^{(1)}| \ll 1$, $\mathcal{P} \approx  |\eta_{p^*}\Theta_{p^*}^{(1)}|^2 = (\eta_{p^*}\alpha)^2$. 
The pulse, if it happens to be that its Fourier coefficients for a tight band of frequencies near $\omega_{p^*}$ dominate in their modulus, would roughly have an average power of $\bar{A} \approx \alpha/\tau$ [see (\ref{eq:TP1})].

\section{Results}
\label{sec:results}

\begin{figure*}[ht!]
\includegraphics[width=\textwidth]{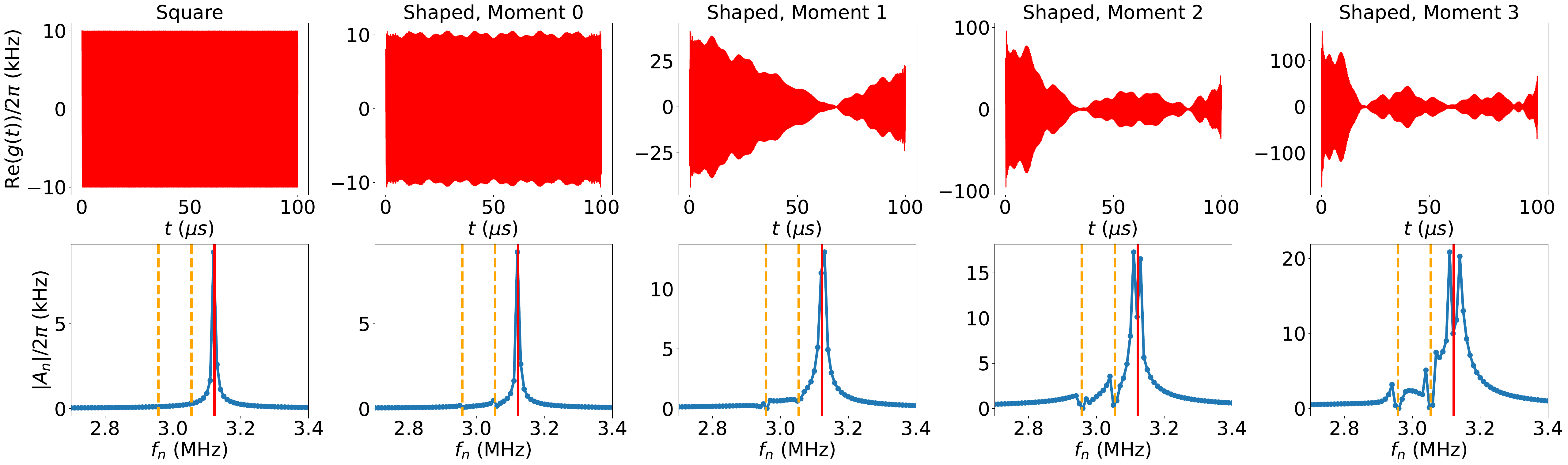}
\caption{Square and shaped pulses with moments of stabilization $K \in \{0,1,2,3\}$, for $N=N'=3$, $\alpha=1$, and $\tau=100\mu $s. 
Top: $\Re[g(t)]$ of each pulse as a function of time $t$. Bottom: Modulus of the Fourier coefficients $|A_n|$ of each pulse as a function of frequency $f_n=n/\tau$. The yellow dashed vertical lines represent the non-target mode frequencies and the red solid line represents the target mode frequency. 
}
\label{fig:pulse_prof}
\end{figure*}

In this section, we demonstrate the viability of our pulse-shaping techniques proposed via numerical simulations of a three-ion chain. To benchmark the performance of our pulse-shaping characterization tools as compared to their traditional square-pulse counterparts, we simulate the bright-state populations $P$ and $\mathcal{P}$ induced by the multi-mode Hamiltonian (\ref{eq:HNmode}) and the single-mode Hamiltonian (\ref{eq:H1mode}), respectively, -- see Appendix A of Ref.~\cite{Mingyu2022} for the simulation implementation detail -- where the initial qubit state is $\ket{0}$ and all motional modes are initially in the ground state. We then use the fractional difference between the induced qubit populations by the two models $\mathcal{E}:=|P-\mathcal{P}|/\mathcal{P}$, hereafter referred frequently as the fractional population error, resulting from the CMC and detuning error, as a proxy for the characterization error. 

It is worth noting that since $P\approx\sin^2(\eta_{p^*}\alpha)$ (see  Appendix~\ref{sec:appendix_population}), $P\approx (\eta_{p^*}\alpha)^2$ in the small-population regime, so the fractional error in $P$ in relation to the fractional error in $\eta_{p^*}$ is given by $\frac{dP}{P} \approx \frac{2d\eta_{p^*}}{\eta_{p^*}}$. Thus, in the presence of fitting errors or experimental imperfections, $\mathcal{E}$ serves as a lower bound to the fractional characterization uncertainty of $\eta_{p^*}$. Computing the proxy error $\mathcal{E}$ for both shaped and square pulses can hence help us reveal the parameter regime in which shaped pulses offer an advantage over square pulses.

Specifically, to see how the parameter regime may be found, we can consider the leading-order errors that remain after pulse shaping and contribute significantly to ${\mathcal E}$, which can be broken down as follows. First, residual CMC errors of order $O(\eta^2)$ or higher remain due to the second- or higher-order Magnus terms. Second, detuning errors arise from higher-order contributions of the first-order Magnus integral $ \frac{\delta_p^\kappa}{\kappa!} \frac{\partial^\kappa}{\partial \omega_p^\kappa}\Theta_{p}^{(1)}$ ($\kappa > K$), where $K$ is the moment of stabilization. Third, additional detuning errors arise from non-target higher-order Magnus integrals. Thus, the dependence of $\mathcal{E}$ to various parameters such as pulse power, pulse duration, and detuning can be predicted based on the residual error sources listed above. Indeed, we aim to identify parameter choices that result in smaller $\mathcal{E}$ for shaped pulses compared to that for square pulses by carefully examining each dependence.

We note in passing that, for a linear chain of three ions indexed by $j\in\{0,1,2\}$ (labeled from left to right) with three modes indexed by $p\in\{0,1,2\}$ (labeled in the ascending order of mode frequencies), we focus throughout this section on ion $j=2$, probing the largest-frequency mode $p^*=2$, as a concrete example; This choice is arbitrary and other ion-mode pairs can be considered straightforwardly in our procedures. 
Further, we use a realistic set of values of $\omega_p$ and $\eta_{j,p}$, reported in Appendix~\ref{app:mode_params}, for our simulated experiments (three-mode Hamiltonian) throughout this section.

Figure~\ref{fig:pulse_prof} shows an example square pulse and shaped pulses with various moments of stabilization $K$, for the aforementioned example choice of the ion and the mode. The pulse components in the frequency domain are mostly concentrated near the target-mode frequency $\omega_{p^*}{=}\omega_2$. The small wiggles around the non-target mode frequencies $\omega_0$ and $\omega_1$ can be interpreted as signatures of active cancellation of the CMC error. Similar features are observed for other choices of ion-mode pairs (not shown). 

The rest of the section is organized as follows.
We first study the $\mathcal{E}$ scaling of square and shaped pulses without stabilization due to the CMC effect in Sec.~\ref{sec:result_cmc}. We then study the $\mathcal{E}$ scaling of square and shaped pulses with various degrees of stabilization in the presence of both errors, due to CMC and detuning, in Sec.~\ref{sec:result_detuning}. Finally, we compare $\mathcal{E}$ between various shaped and square pulses across all parameter regimes in Sec.~\ref{sec:result_final_comparison}.

\subsection{Fractional population error due to CMC}
\label{sec:result_cmc}

We first analyze how $\mathcal{E}$ due to the CMC depends on the Rabi frequency scaling factor $\alpha$ and the pulse length $\tau$. Here the detuning error and stabilization are not considered. We note that intuitively, the CMC error comes from non-zero spectral decomposition of pulse components at non-target mode frequencies, as shown for a square pulse in Fig.~\ref{fig:pulse_prof}. The magnitude of the CMC error due to non-target mode $p$ is then roughly determined by $\bar{A}/|\omega_{p^*} - \omega_p|$, where $\bar{A}$ is the average Rabi frequency of the pulse as defined in Sec.~\ref{sec:method}.

\begin{figure}[ht!]
\includegraphics[width=\columnwidth]{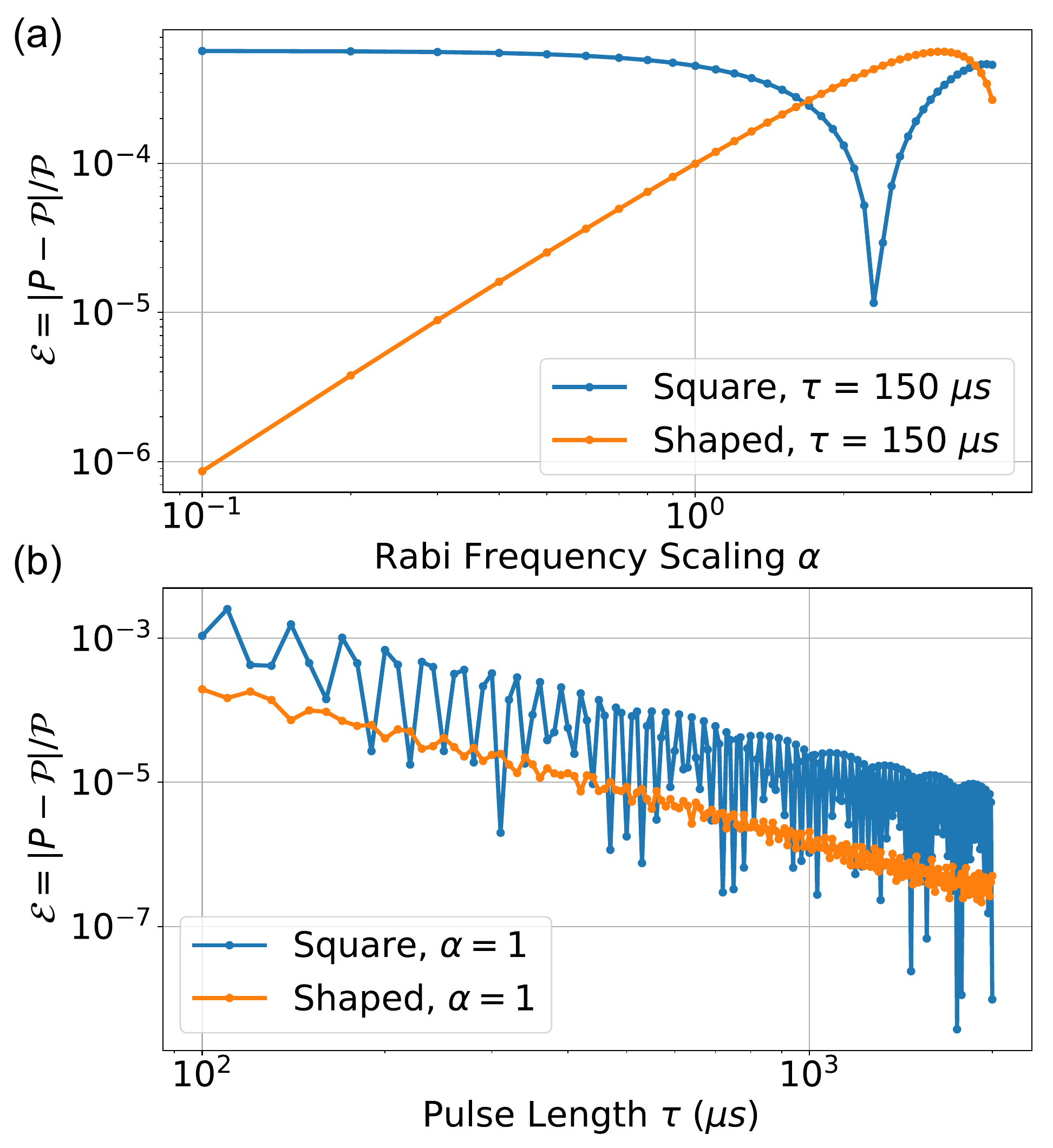}
\caption{Fractional qubit-population error $\mathcal{E}$ for square and shaped (without stabilization) pulses as a function of (a) Rabi frequency scaling $\alpha$ when $\tau=150$$\mu$s and (b) pulse length $\tau$ with the fixed value of $\alpha=1$.}
\label{fig:2}
\end{figure}

\begin{figure*}[ht!]
\includegraphics[width=\textwidth]{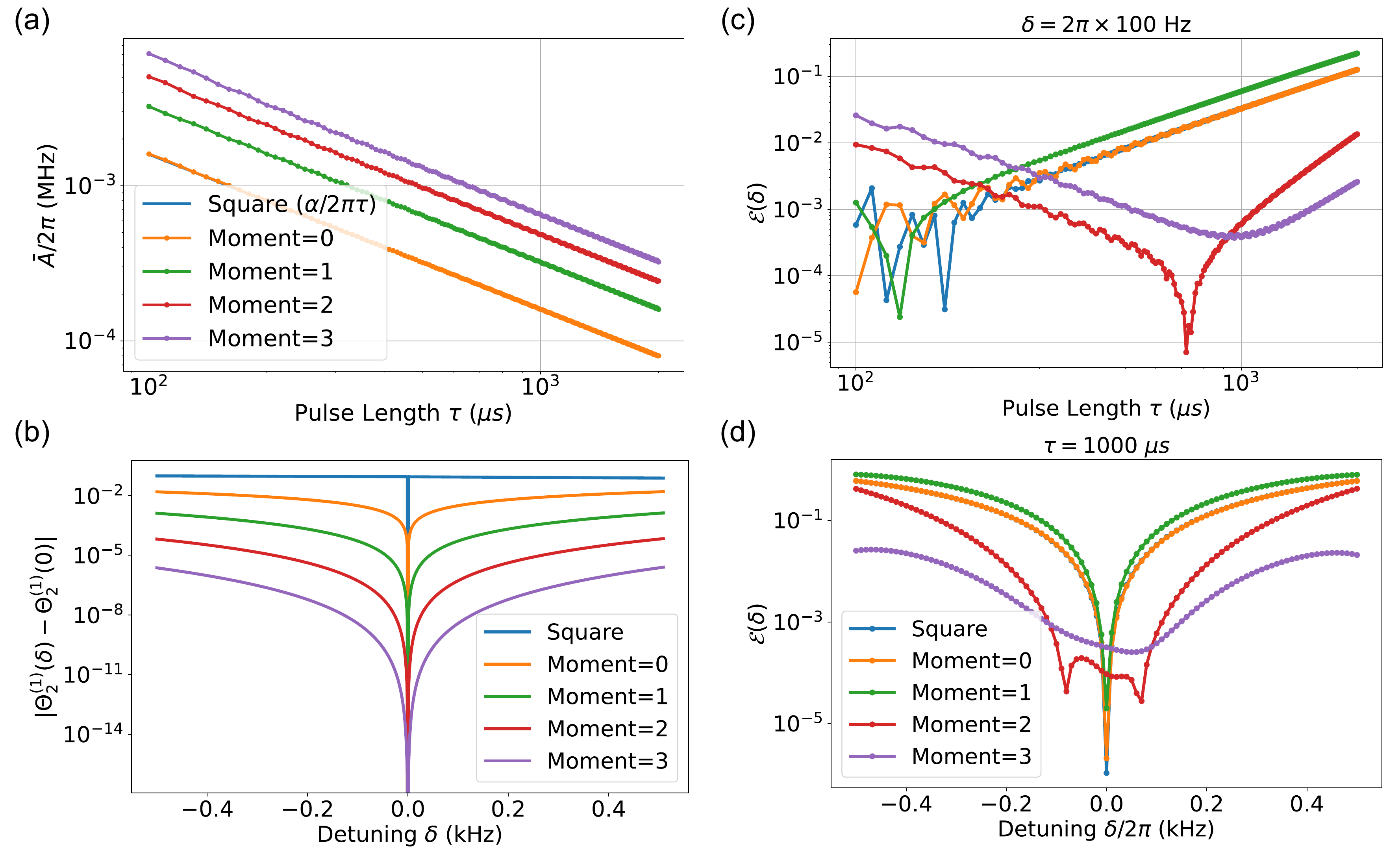}
\caption{Performance metrics of square pulses and shaped pulses with various moments of stabilization. For all pulses, $\alpha=1$. (a) Average Rabi frequencies for various values of gate time $\tau$. The blue line marking $\bar{A}/2\pi=\alpha/2\pi\tau=1/2\pi\tau$ for square pulses (not visible) coincides with the orange line for moment-0 shaped pulses at the scale shown. (b) Deviation in the first-order Magnus integral of target mode $p^*=2$ for various values of mode-frequency detuning $\delta$. (c) Fractional population errors at $\delta=2\pi\times 100$ Hz as a function of pulse length $\tau$. (d) Fractional population errors ${\mathcal E}$ at $\tau = 1000$$\mu$s as a function of detuning $\delta$. The blue curve for square pulses mostly coincides with the orange curve for moment-0 shaped pulses. (c) and (d) share the figure legends of (d).}
\label{fig:3}
\end{figure*}

Figures~\ref{fig:2}(a) and (b) compare the population error $\mathcal{E}$ solely due to the CMC, between square and shaped pulses, as a function of varying $\alpha$ and $\tau$, respectively. 
First, in Fig.~\ref{fig:2}(a), we observe that for the square pulses $g(t) = \bar{A}e^{-i \omega_{p^*} t}$, $\mathcal{E}$ remains constant with increasing $\alpha$ (except for the dip at $\alpha \approx 2.3$ that will be explained later). This is because the major contribution to $\mathcal{P}$ and $|P - \mathcal{P}|$ comes from $|\Theta_{p^*}^{(1)}|^2$ and $|\Theta_{p}^{(1)}|^2$ ($p \neq p^*$), respectively, which are both proportional to $\bar{A}^2 \propto \alpha^2$. Meanwhile, for the shaped pulses, $\mathcal{E}$ is proportional to $\alpha^2$ when $\alpha \lesssim 3$. This is because for shaped pulses, $\Theta_{p}^{(1)} = 0$ for $p \neq p^*$, so the major contribution to \mbox{$|P - \mathcal{P}|$} comes from $|\Theta^{(2)}_{p,p'}|^2$ [$(p,p') \neq (p^*, p^*)$] that is proportional to ${\bar{A}}^4 \propto \alpha^4$; see Appendix~\ref{app:derivations} for details. Due to this difference in the scaling behavior of $\mathcal{E}$ with respect to $\alpha$, shaped pulses are guaranteed to outperform square pulses when $\alpha$ is chosen to be sufficiently small, say, \mbox{$\alpha \lesssim 1$}. 

Figure~\ref{fig:2}(b) shows that when $\alpha{=}1$ is fixed, $\mathcal{E}$ of both square and shaped pulses are roughly proportional to $\tau^{-2}$. This agrees with the observation that (i) \mbox{$\bar{A} \tau \approx \alpha = 1$}, (ii) $|\Theta_{p}^{(1)}|^2$ ($p \neq p^*$) of the square pulse is proportional to $\bar{A}^2 \propto \tau^{-2}$, and (iii) the dominant term of $|\Theta_{p,p'}^{(2)}|^2$ of the shaped pulse (when $\tau \gg |\omega_{p^*} - \omega_{p'}|^{-1}$) is proportional to $(\bar{A}^2 \tau)^2 \propto \tau^{-2}$ (see Appendix~\ref{app:derivations} for details). As $\mathcal{E}$ of the square and shaped pulses have the same scaling behavior with respect to $\tau$, we expect the advantage of shaped pulses when $\alpha$ is small to hold for all pulse lengths. 

We note that, as $\alpha$ and $\tau$ of the square pulses are varied, as evidenced in Fig.~\ref{fig:2}, $\mathcal{E}$ is sharply suppressed at certain ``sweet-spot'' values of $\alpha$ and $\tau$, sometimes by more than an order of magnitude compared to the neighboring values. For Fig.~\ref{fig:2}(a), a likely cause of this dip for the square pulses is the cancellation between the first-order and the higher-order CMC contributions to the qubit population (see Appendix~\ref{sec:appendix_second_order_magnus} for details). For Fig.~\ref{fig:2}(b), the dips occur for the most part when the first-order Magnus integral $\Theta_p^{(1)}$ in \eqref{eq:cmc} evaluates to zero for non-target modes, i.e., $\Theta_p^{(1)}=0$ when $\omega_p-\omega_{p^*}=2\pi l/\tau$ for arbitrary integer $l$. We note that the fact that some of the square pulses have $\mathcal{E}$'s that fall below those of shaped pulses indicates that similar phenomena might be occurring for high-order contributions to CMC error as well. Shaped pulses on the other hand do not have these dips due to the multiple frequency components of the pulse, as the different components are unlikely to be naturally orchestrated to accidentally cancel higher-order Magnus integrals at a given $\tau$.

While it may be desirable to choose $\alpha$ and $\tau$ at these ``sweet spot'' values for our mode characterization, in practice, we cannot perform \textit{a priori} calculations to find the sweet spots, as the mode parameters are yet to be characterized and subject to drifts during the experiment. Thus, actively nulling the first-order contribution of the non-target modes by using a shaped pulse is a more reliable method of suppressing $\mathcal{E}$, and thereby increasing the accuracy of $\eta_p$ estimation.

We further anticipate the shaped pulses to outperform the square pulses in fending off the CMC error for up to about tens of qubits. 
Note, due to the nulling conditions whose number scale linearly in the number of ions, shaped pulses in general demand larger power requirement than square pulses. 
Just as in the gate pulse shaping~\cite{Blumel21}, the extra power demand becomes pronounced when there are insufficient frequency components in the pulse Fourier basis near each mode frequency that needs to be nulled during the CMC suppression.
Since the post-suppression CMC error in the shaped pulse is dominated by the second-order Magnus term that scales quadratically in $\bar{A}$, too large of a power requirement can cause more harm than good.
For $\tau \gtrsim 100\mu$s with mode frequencies within a band of $\sim 0.3$MHz, a typical operating condition of trapped-ion quantum computers today, we can expect a relative abundance of the number of Fourier basis components over the number of mode frequencies within the band for up to tens of ions.
As reported in Appendix~\ref{app:applicability}, we confirm explicitly up to a seven-ion chain that the power requirements remain virtually the same between the square and shaped pulses. 

\subsection{Fractional population errror due to detuning}
\label{sec:result_detuning}

We compare in this section square pulses to shaped pulses with stabilization in the presence of detuning errors. To start, Fig.~\ref{fig:3}(a) shows the average Rabi frequencies $\bar{A}$ required for these pulses to reach $|\Theta_{p^*}^{(1)}|=1$ with various pulse lengths $\tau$. 
We see that for all pulses $\bar{A}$ is inversely proportional to $\tau$. In particular, $\bar{A}$ increases with a larger moment of stabilization. This is because, as seen from the bottom panels of Fig.~\ref{fig:pulse_prof}, pulses with a larger moment of stabilization have larger off-resonant frequency components, which in turn require a larger resonant frequency component to induce the same amount of qubit population inversion, as compared to pulses with smaller off-resonant components. 

We next compare the performance of square and shaped pulses with stabilization as a function of mode-frequency detuning. For simplicity, we consider a uniform detuning model $\omega_p \rightarrow \omega_p + \delta$ for all $p$ (assuming $|\delta|\ll |\omega_p-\omega_{p'}|\;\forall p,p'$), although the advantage of shaped pulses applies for non-uniform detuning as well. Figure~\ref{fig:3}(b) shows the magnitude of the difference between the first-order Magnus integrals with and without the detuning, $|\Theta_{p^*}^{(1)}(\delta)-\Theta_{p^*}^{(1)}(0)|$, as a function of $\delta$. As expected, shaped pulses with a greater moment of stabilization offers a wider detuning range over which $|\Theta_{p^*}^{(1)}(\delta)-\Theta_{p^*}^{(1)}(0)|$ is kept within a small, pre-determined threshold.

To study the effects of the detuning error in conjunction with the CMC error, we show in Figs.~\ref{fig:3}(c) and (d) a generalized version of the fractional population error $\mathcal{E}(\delta)$, defined according to
\begin{align}
    \mathcal{E}(\delta) &:= |P(\delta)-\mathcal{P}(\delta=0)|/\mathcal{P}(\delta=0), \nonumber
\end{align}
where $P(\delta)$ [$\mathcal{P}(\delta)$] is the bright-state population induced by (\ref{eq:HNmode}) [(\ref{eq:H1mode})] under $\omega_p \rightarrow \omega_p + \delta$. 
Specifically for Fig.~\ref{fig:3}(c), we show $\mathcal{E}(\delta)$ as a function of pulse length $\tau$, while fixing the detuning at $2\pi\times 100$Hz. Figure~\ref{fig:3}(d) shows $\mathcal{E}(\delta)$ as a function of $\delta$, while fixing the pulse length at \mbox{$\tau = 860\mu$s}. 

Recall from Fig.~\ref{fig:2}(b) $\mathcal{E}$, in the absence of any detuning and resulting from the CMC, decreased as $\tau^{-2}$ for both square and shaped pulses. When $\delta$ is no longer absent, an erroneous phase that accumulates over the pulse duration $\tau$ emerges, which contributes to $\mathcal{E}(\delta)$ as an increasing function of $\tau$.
Therefore, there is a tug-of-war between the two competing scaling trends in $\tau$, the CMC-induced vs. the detuning-induced. 
There arises a minimum $\mathcal{E}(\delta)$ for some combinations of of $\tau$ and $\delta$.

Figure~\ref{fig:3}(c) shows $\mathcal{E}(\delta)$ as a function of $\tau$, where the transient $\tau$-scaling behavior due to the tug-of-war is pronounced for moment-2 and 3 shaped pulses. 
Indeed, shaped pulses with higher moments of stabilization has a larger CMC error, owing to their larger power requirement (see Fig.~\ref{fig:3}(a)), hence the stronger CMC signature when $\tau$ is small. While not as pronounced in our specific example cases, the transient $\tau$-scaling does appear in lower-moment shaped pulses and square pulses as well (see Appendix~\ref{app:ultimate_comparison}), especially for small $\delta$ or large $\bar{A}$.

A tug-of-war analogy between different errors is also applicable for the observed scaling behavior of ${\mathcal E}(\delta)$ with respect to detuning, shown in Fig.~\ref{fig:3}(d). When $\delta$ is small and CMC error dominates, moment-0 shaped pulses, and in our particular choice of parameters square pulses as well (recall the $\mathcal{E}$ ``dips" for square pulses in Fig.~\ref{fig:2}(b); different pulse lengths change the performance of square pulses dramatically), outperform the rest. This is so, since higher-moment shaped pulses have larger high-order CMC error, as they demand increased power requirement. As $\delta$ increases and detuning error begins to dominate, higher-moment shaped pulses show superior performance as a result of stabilization, with the exception of moment-1 shaped pulses, which we address next. 

In both Figs.~\ref{fig:3}(c) and (d), moment-1 shaped pulses tend to perform the worst in $\mathcal{E}(\delta)$ scaling with respect to both $\tau$ and $\delta$, despite that the first-order derivative of $\Theta_{p^*}^{(1)}$ is nulled as evidenced in Fig.~\ref{fig:3}(b). The discrepancy between our objective quantity $\theta_{p^*}^{\rm{det}} = \Theta_{p^*}^{(1)}(\delta)-\Theta_{p^*}^{(1)}(0)$ to null and $\mathcal{E}(\delta)$ is indeed elaborated in detail in Appendix~\ref{app:even_odd_stabilization}, which explains the poor performance by moment-1 pulses. To summarize, this discrepancy results in better performance of shaped pulses with even-moment stabilization because the leading-order terms of $\mathcal{E}$ have even order dependencies on $\delta$ (quadratic, quartic, etc.). As a result, odd-moment stabilizations of $\theta_{p^*}^{\rm{det}}$ do not have the intended effect of additionally stabilizing $\mathcal{E}(\delta)$, while the stabilizations incur power increase (hence increased CMC contributions from the second-order Magnus terms). In the case of Figs.~\ref{fig:3}(c) and (d), square pulses and moment-0 shaped pulses thus mostly outperform moment-1 shaped pulses, and moment-2 pulses outperform moment-3 pulses in some regimes.

\subsection{Fractional population error comparison between square and shaped pulses}
\label{sec:result_final_comparison}

\begin{figure}[ht!]
\includegraphics[width=\columnwidth]{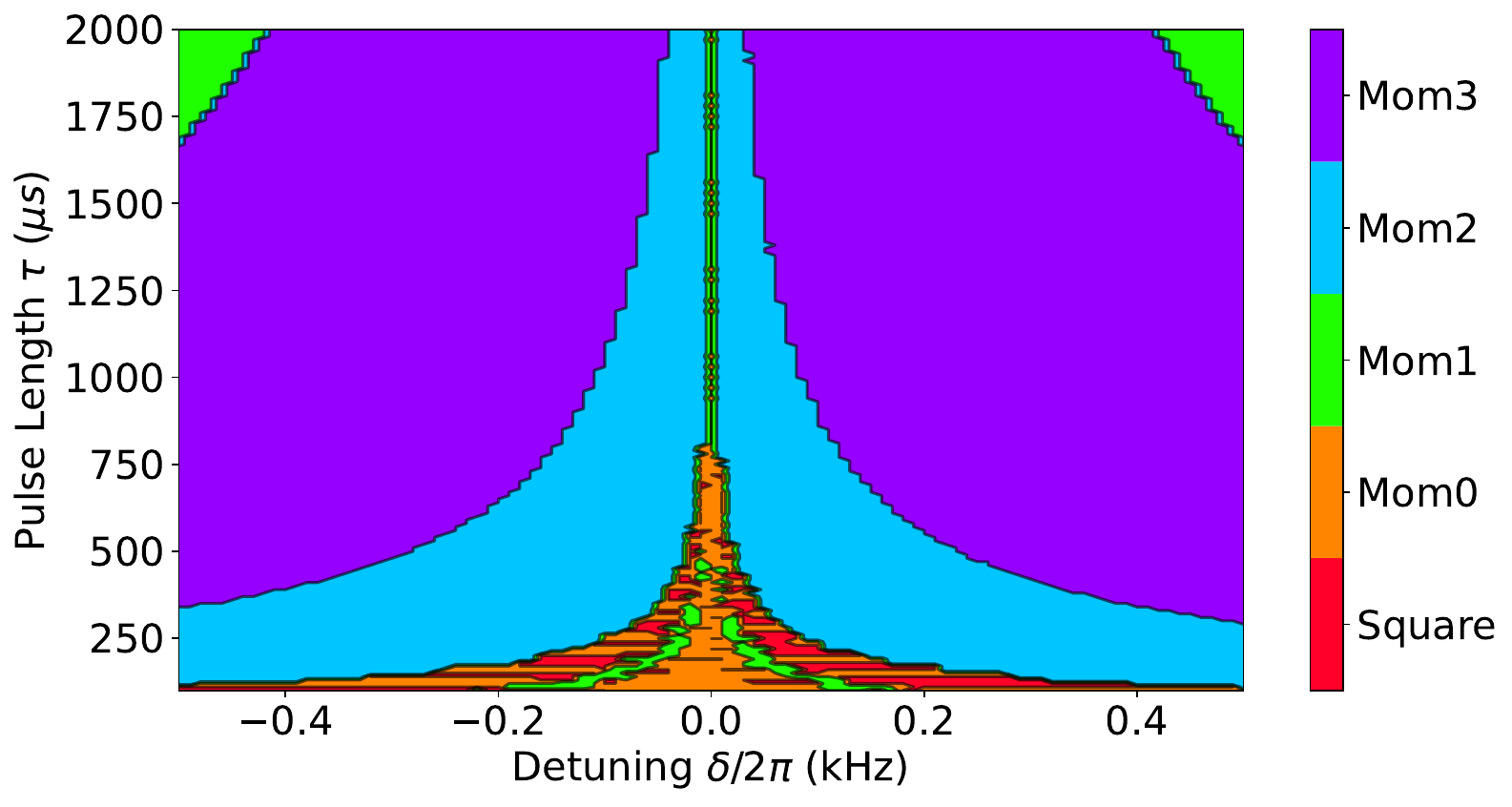}
\caption{Color represents the pulse that achieves the smallest fractional qubit-population error $\mathcal{E}(\delta)$, where the error is due to both the CMC and detuning, for various values of pulse duration $\tau$ and mode-frequency detuning $\delta$. For each pulse duration, the candidates of pulses are square pulse and shaped pulses with moments of stabilization 0 to 3. For all pulses, $\alpha=1$.
}
\label{fig:4}
\end{figure}

Recall our goal is to identify the optimal choice of pulse parameters, such as square vs. shaped, moments of stabilization, or duration, such that the mode characterization for a given system, specified by its rough mode spacings and detuning stability, is achieved with the lowest error.
To this end, Fig.~\ref{fig:4} shows which pulses out of square and moments-$K$ shaped pulses, $K\in\{0,1,2,3\}$, perform the best in terms of the fractional population errors $\mathcal{E}(\delta)$, in the presence of both CMC and detuning errors for various values of pulse length $\tau$ and detuning $\delta$. As expected, shaped pulses outperform square pulses in most parameter regimes, where higher-moment shaped pulses tend to perform best for long pulse lengths unless the detuning error is minuscule ($|\delta| \ll \alpha/\tau$) and lower-moment pulses tend to perform best for short pulse lengths over a sizable range of detuning errors. 

We refer the readers to Appendix~\ref{app:ultimate_comparison} for a full comparison of $\mathcal{E}(\delta)$ between all of the pulses considered in Fig.~\ref{fig:4}, beyond just the best performing one, for various values of $\tau$ and $\delta$. In summary, we observe that, for a moderate amount of mode-frequency uncertainty $|\delta| \lesssim 2\pi \times 80$Hz and mode spacings $\omega_2-\omega_1 = 2\pi \times 68.0$kHz and $\omega_1-\omega_0 = 2\pi \times 96.8$kHz, moment-2 pulses of length $\tau \geq 1000\mu$s achieve the smallest population error ($10^{-4} \lesssim \mathcal{E}(\delta) \lesssim 10^{-3}$) among all the pulses considered across all pulse lengths $\tau$, for the case of three ions and three modes.
Our detailed findings reported in Appendix~\ref{app:ultimate_comparison} are in line with the scaling trends summarized in Figs.~\ref{fig:2} and \ref{fig:3}, making an optimal choice of pulse parameters based on the uncertainty in the mode-frequency estimation (which results in unwanted detuning) and the available pulse power/duration possible, at least roughly. We expect similar scaling behaviors to hold for larger numbers of ions, which our pulse shaping tool can be readily applied to (see Appendix~\ref{app:applicability}).

\section{Discussion}
\label{sec:discussion}
\subsection{Efficiency of mode characterization}
\label{subsec:efficiency}

As much as accuracy is important in mode characterization, its efficiency also needs to be considered~\cite{Mingyu2022}. In an experimental setup, the parameters to be characterized inevitably drift over time. Therefore, quickly and frequently updating the parameters is necessary for efficient and accurate operation of a quantum computer~\cite{Maksymov2022}. However, in order to estimate $\eta_{j,p}$ with low uncertainty, the BSB transition needs to be performed to a sufficiently large degree (by using longer pulse length $\tau$ or larger average Rabi frequency $\bar{A}$), and this needs to be repeated for sufficiently many shots. In particular, the requirement for large number of shots is a major bottleneck for efficient mode characterization, as for each shot, cooling, state preparation, and measurement need to be performed in addition to the BSB transition~\cite{Mingyu2022}. 

As an example, we consider estimating $\eta_{p^*}$ with uncertainty $\Delta \eta$. Then, the difference in qubit population due to the change in $\eta_{p^*}$ by $\Delta \eta$ should exceed the shot noise; thus, the estimation is successful only if
\begin{equation}
    \frac{d \mathcal{P}}{d \eta_{p^*}} \Delta \eta \geq \sqrt{\frac{\mathcal{P} (1-\mathcal{P})}{S}},
\end{equation}
where $S$ is the number of shots. 
As briefly mentioned in Sec.~\ref{sec:method}, in the limit of small $|\eta_{p^*} \Theta^{(1)}_{p^*}| = |\eta_{p^*}| \alpha$, $\mathcal{P} \approx \left(\eta_{p^*} \alpha \right)^2$, and thus, up to a leading order in $\eta_{p^*} \alpha$, we obtain
\begin{equation}
    S \geq \left(2 \alpha \Delta \eta\right)^{-2}.
\end{equation}
Therefore, in order to achieve uncertainty $\Delta \eta$ with fewer shots, it is desirable to use a pulse of larger $\alpha$, which requires larger pulse power or a longer pulse length. We remind the readers that $\alpha$ is fixed to 1 for the analysis in Figs.~\ref{fig:2}(b) and \ref{fig:3}.

The limitation of the proposed pulse-shaping method is that the shaped pulses are guaranteed to achieve smaller errors than the square pulses only in the regime of $\alpha \lesssim 1$, as shown for the CMC error in Fig.~\ref{fig:2}(a). This is because for larger $\alpha$, the contribution from the second- and higher-order Magnus terms to the qubit population dominates that from the first-order Magnus terms that can be silenced by pulse shaping. Therefore, the applicability of our pulse-shaping method is limited to the case where performing a large number of shots is feasible and the uncertainty in the mode-parameter estimation due to the shot noise is dominated by the inaccuracy due to the CMC and detuning errors. 

A possible solution one can think of is to further silence the effect of the second-order Magnus terms by pulse shaping. A method for nulling second-order Magnus terms is proposed in Appendix~\ref{app:nulling_second_magnus}. However, based on the method, it is not feasible to silence all the second-order Magnus terms due to the positive-definiteness of the nulling condition matrices describing $\Theta_{p,p'}^{(2)}$ for $p=p'$ (details in Appendix~\ref{app:nulling_second_magnus}). 

Furthermore, we leave to future research whether removing the CMC and detuning errors by pulse shaping can be done without the Lamb-Dicke approximation ($|\eta_p| \ll 1$), similar to the works on pulse shaping for ultrafast two-qubit gates~\cite{GarciaRipoll03, Steane14, WongCampos17, Schafer18}. This potentially retains the advantage of pulse shaping for larger $\alpha$ and thus for an efficient mode characterization with fewer shots.

The efficiency of mode characterization can be further improved by employing parallelization; that is, probing $N$ values of $\eta_{j,p}$ simultaneously by driving BSB transition on all $N$ ions, each ion assigned to a different mode~\cite{Mingyu2022}. This can reduce the number of operations required to characterize all $N \times N'$ values of $\eta_{j,p}$ from $O(N^2)$ to $O(N)$. While na\"ively using the shaped pulses proposed in Sec.~\ref{sec:method} in parallel successfully removes the CMC errors up to first order in each $\eta_{j,p}$, various higher-order CMC errors that are different from those analyzed above may arise. This is because the off-resonant contribution of the BSB transition that probes mode $p_1$ to the population of mode $p_2 \neq p_1$ may interfere with the resonant contribution of the BSB transition that probes mode $p_2$ in parallel. The resulting errors in qubit populations, as well as a method of removing such errors by modifying the pulse-shaping scheme, need to be considered when parallelization is employed in future work. 

\subsection{Additional sources of errors}
\label{subsec:additionalerrors}

In this paper, we identify the CMC and the detuning of mode frequencies as sources of errors in the mode characterization and remove their effects by pulse shaping. Yet there are several other sources of errors that need to be addressed for performing accurate mode characterization. 

First, while we assume that the motional mode is initially at the ground state, in practice mode $p$ can initially be in a state of phonon number $n_{p} \geq 1$ with small but nonzero probability due to imperfect cooling. This is important for characterizing $\eta_p$ using BSB transitions, as the sideband Rabi frequency is proportional to $\eta_p \sqrt{n_p + 1}$. If the temperature, or the average phonon number $\bar{n}_p$, of the motional state after cooling is known, the qubit population can be computed by taking a weighted average over the distribution of initial phonon number $n_p$, where the distribution is straightforward to obtain from $\bar{n}_p$ assuming thermal distribution. The uncertainty in $\bar{n}_p$ may limit the accuracy of $\eta_p$ estimation. Nonetheless, using shaped pulses does not pose additional difficulties compared to using square pulses.

Second, the qubit population is also related to $\eta_p$'s by the Debye-Waller effect, as briefly explained in Sec.~\ref{sec:problem}. In the context of trapped ions, this effect refers to reduction in the Rabi frequency due to the nonzero width of the ion's position wavepacket, which spreads more as the ion is more strongly coupled to the motional modes~\cite{Wineland79, NIST_Bible}. While this effect is at most second order in $\eta_{p}$ and therefore ignored in the linearized Hamiltonian in (\ref{eq:HNmode}) and (\ref{eq:H1mode}), it may have a significant impact on the qubit population especially in long ion chains, as all $N'$ modes contribute to this Rabi-frequency reduction. Thus, the qubit population after the BSB transition with respect to mode $p^*$ depends on the values of all $\eta_{p}$'s ($p=1,\ldots, N'$). 

To estimate $\eta_{p^*}$ accurately, the reduction factors from all $N'$ modes need to be included in the model that predicts the qubit population. However, fitting the measured qubit populations to the model's predictions with all $\eta_p$'s as fitting parameters takes a prohibitively long computational time. Therefore, an iterative fitting routine, where the values of $\eta_p$'s obtained from the previous iterations of fitting are used in the current iteration for estimating $\eta_{p^*}$, is necessary. The algorithm for this fitting routine can be found in Appendix B of Ref.~\cite{Mingyu2022}, which is designed for square pulses but can be straightforwardly extended to shaped pulses as well. Moreover, to briefly mention, some of the methods developed in Ref.~\cite{Mingyu2022} to characterize the sign of the $\eta_p$'s are compatible with our proposed pulse shaping scheme.

Finally, there are various other experimental imperfections that may cause errors in $\eta_p$ characterization. Examples of such imperfections are miscalibration of the pulse parameters, anharmonicity of the motional modes, optical crosstalk, and fluctuations of the mode parameters over time. The magnitudes of characterization errors as a result of the above-listed experimental imperfections need to be taken into account when determining the target uncertainty $\Delta \eta$ in estimating $\eta_p$. The target uncertainty $\Delta \eta$ may then be used for determining various parameters of the mode-characterization protocol, such as $\alpha$, $\tau$, and $S$, aiming for as efficient mode characterization as possible; see Sec. 6.1 of Ref.~\cite{Mingyu2022} for a detailed discussion that applies to both square and shaped pulses.

\section{Outlook}
\label{sec:conclusion}

We devised a pulse-shaping scheme for high-accuracy characterization of the motional-mode parameters of trapped-ion quantum computers. We have extensively tested the effectiveness of our scheme in minimizing the CMC and detuning errors via simulations on a three-ion chain. Our results showed an improvement in the accuracy of $\eta_{j,p}$ estimation, even when the mode frequencies were not well known. 

Our pulse-shaping scheme helps address one of the many engineering challenges for building large trapped-ion quantum computers. As the number of qubits within a trap increases, an accurate and efficient mode characterization in the presence of other motional modes becomes increasingly challenging.
By leveraging the degrees of freedom in pulse design for these experiments, our methodology offers one potential path forward to overcoming this bottleneck, namely, accurately probing mode parameters in the presence of model violations and experimental imperfections. One of the exciting future directions may be to explore parallel pulse shaping, where multiple ions are addressed simultaneously to characterize mode parameters more efficiently.

The accurate and efficient characterization of these mode parameters holds the key to high-fidelity trapped-ion operations, such as parallel entangling gates \cite{Bentley20, Grzesiak20}. Briefly, these gates are implemented by illuminating each ion by custom-designed pulse shapes such that the induced interactions between the internal states of the ion qubits match those intended as quantum gates acting over multiple pairs of qubits simultaneously. Such interactions are, just as in a single two-qubit gate between two ion qubits, mediated by motional modes of the ion chain. As such, the parallel entangling gates intricately depends on the precise knowledge of the mode parameters for their high-fidelity implementation. Provided that these parallel gates have been proven to be versatile in enabling efficient quantum computing over a wide class of quantum programs~\cite{grzesiak2022efficient,bravyi2022constant}, the improvement achieved in characterizing mode parameters enabled by our methodology is expected to play a crucial role in realizing efficient quantum computation.

Our methodology of applying pulse shaping and other optimal control techniques for characterization tasks may readily be adopted to be applicable for different quantum computer architectures as well. Systems that rely on spectroscopic system parameter characterization like trapped ions can likely benefit from pulse shaping in their characterization by effectively eliminating significant model violations. We hope our work, which opens a large trade space of system-level optimization for exploration and exploitation, will help boost the effort of building a large-scale quantum computer.

\section*{Acknowledgements}
\label{sec:ack}
Q.L. and M.K. thank Kenneth R. Brown for helpful suggestions. M.K. was supported by the Army Research Office, W911NF-21-1-0005.

\bibliography{bib}

\onecolumngrid
\appendix
\label{sec:appendix}
\section{Values of mode parameters}
\label{app:mode_params}
The motional-mode frequencies and Lamb-Dicke parameters used in the simulations of Sec.~\ref{sec:results} are given in Table \ref{tab:valuesN3}. The values are obtained from a mode-structure model of equidistantly spaced ions trapped in an HOA2.0 trap~\cite{HOA}. 

\begin{table*}[ht]
\renewcommand*{\arraystretch}{1.5}
\begin{tabular}{ c | c | c | c }
\hline
& $p = 0$ & $p = 1$ & $p = 2$\\
\hline
\begin{tabular}{@{}c@{}}$\omega_p / 2\pi$ \\ (MHz)\end{tabular} & 
2.9574 & 3.0542 & 3.1222 \\
\hline
$\eta_{j=0,p}$ & $-0.0457$ & $0.0776$ & $0.0625$\\
 $\eta_{j=1,p}$ & $0.0909$ & $-2.77 \times 10^{-6}$ & $0.0629$ \\
$\eta_{j=2,p}$ & $-0.0457$ & $-0.0776$ & $0.0625$ \\
\hline
\end{tabular}
\caption{Values of mode parameters for $N = N' = 3$.}\label{tab:valuesN3}
\end{table*}

\section{Magnus Term Derivations}
\label{app:derivations}

In this section, we derive the Magnus terms and analyze their scaling trends with respect to parameters such as the average Rabi frequency $\bar{A}$, pulse length $\tau$, and detuning $\delta_p$ of the mode frequency. We start with (\ref{eq:HI}), the interaction Hamiltonian when the $j$-th ion is illuminated by a laser of pulse $g_j(t)$. As we consider the case where only one ion is illuminated at a time, we omit index $j$ for brevity. 

In order to analyze high-order terms and their scaling trends with respect to the parameters of our interest, we keep up to the second-order Taylor expansion term in expanding the exponential within (\ref{eq:HI})
\begin{align}
    \exp{i\sum_p \eta_{p} (\hat{a}_p e^{-i\omega_p t}+\hat{a}_p^{\dagger} e^{i\omega_p t})} \approx \hat{X}_0+\hat{X}_1+\hat{X}_2,
\end{align}
where $\eta_{p} := \eta_{j,p}$, and
\begin{align}
    &\hat{X}_0 = \hat{I},\\
    &\hat{X}_1(t) = i\sum_p \eta_{p}(\hat{a}_p e^{-i\omega_p t}+\hat{a}_p^{\dagger} e^{i\omega_p t}),\\
    &\hat{X}_2(t) = -\frac{1}{2}\left[
    \sum_p \eta_{p} (\hat{a}_p e^{-i\omega_p t}+\hat{a}_p^{\dagger} e^{i\omega_p t})
    \right]^2
\end{align}
are the zeroth-, first-, and second-order Taylor expansion terms in $\eta_{p}$. Here, $\hat{a}_p$ ($\hat{a}^{\dagger}_p$) is the annihilation (creation) operator on mode $p$. We neglect higher-order terms $\hat{X}_n(t)$ in the Taylor series expansion assuming either of the following two situations. For terms that have the same frequencies as the target mode, we can treat them as the Debye-Waller terms that have been discussed in detail in Sec.~\ref{sec:problem} of the main text. For terms that have non-zero frequency differences from the target mode frequencies, we assume that the norm of the corresponding Hamiltonian terms are much smaller than these frequency differences.

We then write out the approximated full unitary corresponding to the Hamiltonian (\ref{eq:HI}) through Magnus expansion, keeping up to the second-order term 
\begin{align}
\label{eq:ap_uni}
\hat{U} \approx \exp{\hat{\Omega}_1+\hat{\Omega}_2}, \nonumber
\end{align}
where
\begin{align}
    \hat{\Omega}_1 &= -i\int_0^\tau dt \hat{H}_I(t) 
    \approx -i \hat{\sigma}^+ \int_0^{\tau}g(t) \left(1+\hat{X}_1(t)+\hat{X}_2(t) \right)dt - h.c. \\
    \hat{\Omega}_2 &= -\frac{1}{2}\int_0^\tau dt_1 \int_{0}^{t_1} dt_2 \left[ \hat{H}_I(t_1), \: \hat{H}_I(t_2) \right] \nonumber \\
    &\approx -\frac{1}{2}\int_0^\tau dt_1 \int_{0}^{t_1} dt_2 \left[ \hat{\sigma}^+ g(t_1)(1+\hat{X}_1(t_1)+\hat{X}_2(t_1))+h.c., \quad \hat{\sigma}^+ g(t_2)(1+\hat{X}_1(t_2)+\hat{X}_2(t_2))+h.c. \right]
\end{align}
are the first- and second-order Magnus terms. As a reminder, $\hat{\sigma}^+$ ($\hat{\sigma}^-$) is the raising (lowering) operator on the qubit. 

Here we assume the pulse to be in the form $g(t)=\sum_n A_n e^{-i\omega_n t}$ over a duration of pulse length $\tau$, where $\omega_n=\frac{2\pi n}{\tau}$ is the frequency of the pulse component with Rabi frequency $A_n$. We consider pulses with low power, i.e., $|A_n| \ll \omega_p, \omega_n$, such that perturbative equations apply. We are interested in the scaling trends of the qubit population and its error with respect to parameters such as $\alpha$, $\tau$, and $\delta_p$. Thus, we analyze the contributions to the qubit population from the Taylor expansion terms up to second order ($\hat{X}_0$, $\hat{X}_1$, and $\hat{X}_2$). We define
\begin{align}
    \label{eq:b7}
    \hat{\Omega}_1^{(r)} &= -i \hat{\sigma}^+ \int_0^\tau g(t) \hat{X}_r(t) dt - h.c.,\\
    \label{eq:b8}
    \hat{\Omega}_2^{(r,s)} &= -\frac{1}{2} \int_0^\tau dt_1 \int_{0}^{t_1} dt_2
    \left[ \hat{\sigma}^+ g(t_1) \hat{X}_r(t_1) + h.c., \:\: 
    \hat{\sigma}^+ g(t_2) \hat{X}_s(t_2) + h.c., \right],
\end{align}
such that $\hat{\Omega}_1 \approx \sum_{r=0}^2 \hat{\Omega}_1^{(r)}$ and $\hat{\Omega}_2 \approx \sum_{r=0}^2 \sum_{s=0}^2 \hat{\Omega}_2^{(r,s)}$. In our following treatment of $\hat{\Omega}_1$ and $\hat{\Omega}_2$, we assume that the norm of some of the terms $\hat{\Omega}_1^{(r)}$ and $\hat{\Omega}_2^{(r,s)}$ are small in comparison to the differences between the frequencies of those terms and the target mode frequency. As a result, we apply the RWA to eliminate them within the Magnus integrals. The specific terms that we rotate away are described in the subsequent sections.

\subsection{First-order Magnus terms}
\label{sec:appendix_first_order_magnus}
Here we evaluate the first-order Magnus term $\hat{\Omega}_1$. We consider pulses with low power ($|A_n| \ll \omega_p, \omega_n$) that are near-resonant to the blue-sideband frequencies; that is, $|A_n|$ is non-negligible only for $n$ such that $|\omega_p - \omega_n| \ll \omega_p$ for some $p$. Then, we obtain 
\begin{align}
    \int_0^\tau g(t)dt \approx 0 \quad &\Rightarrow \quad \hat{\Omega}_1^{(0)} \approx 0, \nonumber\\
    \int_0^\tau g(t) \hat{X}_2(t)dt \approx 0 \quad &\Rightarrow \quad \hat{\Omega}_1^{(2)} \approx 0, \nonumber
\end{align}
by the rotating wave approximation (RWA) described previously following \eqref{eq:b7} and \eqref{eq:b8}. Thus, $\hat{\Omega}_1 \approx \hat{\Omega}_1^{(1)}$, where (taking into account that small detuning $\omega_p \rightarrow \omega_p + \delta_p$ may occur to the mode frequency) $\hat{\Omega}_1^{(1)}$ is given by
\begin{align} \label{eq:Omega_1_1}
    \hat{\Omega}_1^{(1)} &= -i \hat{\sigma}^+ \int_0^\tau g(t) \hat{X}_1(t)dt - h.c. \nonumber \\
    &= \hat{\sigma}^+ \sum_p \eta_p \sum_n  A_n \int_0^\tau dt \left(\hat{a}_p e^{-i (\omega_p + \omega_n + \delta_p) t} + \hat{a}_p^\dagger e^{i (\omega_p - \omega_n + \delta_p) t} \right) - h.c. \nonumber \\
    &\approx \hat{\sigma}^+ \sum_p \eta_p \hat{a}_p^\dagger
    \sum_n  A_n \int_0^\tau dt e^{i (\omega_p - \omega_n + \delta_p) t} - h.c. \nonumber \\
    &= \sum_{p} \eta_p \Theta_{p}^{(1)} \hat{\sigma}^+ \hat{a}_p^{\dagger} - h.c.
\end{align}
In arriving at \eqref{eq:Omega_1_1}, we used the RWA from $|\eta_p A_n| \ll |\omega_p + \omega_n + \delta_p|$ in between line 2 and 3 above and defined $\Theta_p^{(1)} := \sum_n \Theta^{(1)}_{p,n}$, where
\begin{equation} \label{eq:theta_pn_1}
    \Theta^{(1)}_{p,n} := A_n \int_0^\tau dt e^{i(\omega_p - \omega_n + \delta_p) t} 
    = 2 A_n e^{i \omega_{p,n} \tau / 2} \frac{\sin(\omega_{p,n} \tau/2)}{\omega_{p,n}},
\end{equation}
and $\omega_{p,n}:=\omega_p-\omega_n+\delta_p$.

We discuss how $|\Theta_p^{(1)}|$ scales with various parameters. First, we consider the average Rabi frequency $\bar{A} := \sqrt{\sum_n |A_n|^2}$. From (\ref{eq:theta_pn_1}), $\alpha := |\Theta_{p^*}^{(1)}|$ is proportional to the average Rabi frequency $\bar{A}$. Indeed, when obtaining the pulse solution, $\alpha$ is the \textit{Rabi-frequency scaling factor} that is multiplied uniformly to the components $A_n$ of the pulse solution that is normalized such that $|\Theta_{p^*}^{(1)}|=1$; see (\ref{eq:normalizepulse}). For $p' \neq p^*$, $|\Theta_{p'}^{(1)}|$ is proportional to $\bar{A}$ as well.

Next, we consider the pulse length $\tau$. For the case of perfectly resonant square pulse $g(t) = \bar{A} \exp\{-i\omega_{p^*}t + i\phi\}$, we may take the limit $\omega_{p^*,n}\tau \rightarrow 0$ in (\ref{eq:theta_pn_1}) and obtain $|\Theta_{p^*}^{(1)}| =  \bar{A}\tau \propto \tau$. For shaped pulses, taking the limit $\omega_{p^*, n} \tau \rightarrow 0$ is not applicable for all $n$ with non-negligible $|A_n|$; however, we expect $|\Theta_{p^*}^{(1)}| \approx \bar{A} \tau$ for shaped pulses with similar spectrum of $|A_n|$ to square pulse (see moment-0 pulse of Fig.~\ref{fig:pulse_prof} for example). For $p' \neq p^*$, if we take the same limit $\omega_{p^*,n}\tau \rightarrow 0$ in (\ref{eq:theta_pn_1}), then $\omega_{p',n} \rightarrow \omega_{p'} - \omega_{p^*}$, so $|\Theta_{p'}^{(1)}|$ oscillates with angular frequency $(\omega_{p'} - \omega_{p^*})/2$ as $\tau$ is increased (unless $|\Theta_{p'}^{(1)}|$ is actively suppressed to zero by pulse shaping). This is shown by the dips in Fig.~\ref{fig:2}(b) for square pulses.

Finally, we consider the detuning $\delta_p$. Assuming small detuning $|\delta_p| \ll 1/\tau$, Taylor expansion of (\ref{eq:theta_pn_1}) with respect to $\delta_p$ includes non-zero terms for all orders of $\delta_p$. Thus, for a pulse with $K$-th order stabilization, the leading-order contribution of $\delta_p$ to $|\Theta_p^{(1)}|$ is proportional to $\delta_p^{K+1}$, where $K=0$ applies to square pulse or shaped pulse with no stabilization. 

\subsection{Second-order Magnus terms}
\label{sec:appendix_second_order_magnus}
Here we evaluate the second-order Magnus term $\hat{\Omega}_2$. Similarly to the case of $\hat{\Omega}_1$, we find that $\hat{\Omega}_2^{(r,s)}$ is negligibly small for $(r,s) = (0,0)$, $(0,1)$, $(1,0)$, $(0,2)$, and $(2,0)$ by the the same RWA as described after \eqref{eq:b7} and \eqref{eq:b8}. Thus, up to leading order, $\hat{\Omega}_2 \approx \hat{\Omega}_2^{(1,1)}$, where, in the presence of detuning $\omega_p \rightarrow \omega_p + \delta_p$,
\begin{align}
    \hat{\Omega}_2^{(1,1)}&= -\frac{1}{2} \int_0^\tau dt_1\int_0^{t_1}dt_2 
    \Big[ i \hat{\sigma}^+ g(t_1) \sum_p \eta_p (\hat{a}_p e^{-i (\omega_p + \delta_p) t_1} + \hat{a}_p^\dagger e^{i (\omega_p + \delta_p) t_1}) + h.c., \nonumber \\
    &\quad\quad\quad\quad\quad\quad\quad\quad\quad\quad
    i \hat{\sigma}^+ g(t_2) \sum_{p'} \eta_{p'} (\hat{a}_{p'} e^{-i (\omega_{p'} + \delta_{p'}) t_2} + \hat{a}_{p'}^\dagger e^{i (\omega_{p'} + \delta_{p'}) t_2}) + h.c. \Big] \nonumber\\
    &= \frac{1}{2} \sum_{p,p'} \eta_{p} \eta_{p'} \sum_{n,n'} A_n A_{n'}^*
    \int_0^\tau dt_1 \int_0^{t_1} dt_2 e^{i\omega_{p,n}t_1} e^{-i\omega_{p',n'}t_2} (\hat{\sigma}^z \hat{a}_p^{\dagger}\hat{a}_{p'}+\hat{\sigma}^-\hat{\sigma}^+ \delta_{p,p'}) - h.c. \nonumber\\
    &=\sum_{p,p'}\eta_{p}\eta_{p'} \Theta_{p,p'}^{(2)} (\hat{\sigma}^z \hat{a}_p^{\dagger}\hat{a}_{p'}+\hat{\sigma}^-\hat{\sigma}^+ \delta_{p,p'}) - h.c. 
    \label{eq:sec_order_term}
\end{align}
Here, $\hat{\sigma}^z$ is the Pauli-Z operator on the qubit, $\delta_{p,p'}$ is the Kronecker delta symbol, and $\Theta_{p,p'}^{(2)} := \sum_{n,n'} \Theta_{p,p',n,n'}^{(2)}$ where
\begin{align} \label{eq:theta_ppnn_2}
    \Theta_{p,p',n,n'}^{(2)} &:=  A_n A_{n'}^*
    \int_0^\tau dt_1 \int_0^{t_1} dt_2 e^{i\omega_{p,n}t_1} e^{-i\omega_{p',n'}t_2} \nonumber \\
    &= \frac{A_n A_{n'}^*}{2 \omega_{p,n}\omega_{p',n'}(\omega_{p,n}-\omega_{p',n'})} \left(\omega_{p',n'}-\omega_{p,n}e^{i(\omega_{p,n}-\omega_{p',n'})\tau}+(\omega_{p,n}-\omega_{p',n'})e^{i\omega_{p,n}\tau} \right). 
\end{align}

With all other parameters fixed, $|\Theta_{p,p'}^{(2)}|$ is proportional to $\bar{A}^2$ as $\Theta_{p,p',n,n'}^{(2)} \propto A_n A^*_{n'}$. This $\Theta_{p,p'}^{(2)}$ [$(p,p')\neq(p^*,p^*)$] is the second-order contribution to the CMC error for square pulses. Note that the first-order contribution $\Theta_{p'}^{(1)}$ ($p'\neq p^*$) is proportional to $\bar{A}$. This explains the dip in the population-error curve for square pulses in Fig.~\ref{fig:2}(a). As $\alpha \propto \bar{A}$ is increased from a small value (with fixed $\tau$), $|\Theta_{p,p'}^{(2)}|$ is initially smaller than $|\Theta_{p'}^{(1)}|$ but grows faster. At the point where the magnitude of $|\Theta_{p,p'}^{(2)}|$ roughly matches $|\Theta_{p'}^{(1)}|$, the effects of first- and second-order CMC to the population error may cancel each other out, which, as we pointed out in the main text, is a likely cause of the dip for square pulses in Fig.~\ref{fig:2}(a). On the other hand, for shaped pulses, $\Theta_{p,p'}^{(2)}$ is the leading-order contribution to the CMC error, as $|\Theta_{p'}^{(1)}|$ is suppressed to zero for all $p' \neq p^*$.

The scaling behavior of $|\Theta_{p,p'}^{(2)}|$ with respect to $\tau$ is rather complicated. We find three categories of mode pairs $(p,p')$ such that the pair(s) in the same category exhibit similar scaling behavior: (i) $(p=p^*, p'\neq p^*)$ or $(p \neq p^*, p'=p^*)$ or $(p, p'=p\neq p^*)$, (ii) $(p=p^*, p'=p^*)$, and (iii) $(p\neq p^*, p'\neq p^*)$ where $p \neq p'$. 

First, consider the case where $p=p^*$ and $p' \neq p^*$. Similarly to the first-order Magnus integral above, we may take the limit $\omega_{p^*,n}\tau \rightarrow 0$, which is valid for square pulse and shaped pulses of similar spectrum to square pulse. This leads to
\begin{equation} \label{eq:theta_ppnn_2_lim}
   \Theta_{p^*,p',n,n'}^{(2)} \rightarrow i A_n A^*_{n'} \left( 
    \frac{\tau}{2 \omega_{p',n'}} - e^{-i \omega_{p',n'}\tau/2} 
    \frac{\sin(\omega_{p',n'}\tau/2)}{\omega_{p',n'}^2}
   \right).
\end{equation}
It is reasonable to consider pulse lengths that are much longer than the inverse of the \textit{spacing} between the mode frequencies (albeit much shorter than the inverse of the mode-frequency value). In such case, for the types of pulses considered here, $\tau \gg |\omega_{p',n'}|^{-1}$ for all $n'$ such that $|A_{n'}|$ is non-negligible. Then, the first term, which is proportional to $\tau$, dominates the second term. Thus, for the types of pulses and the regime of pulse lengths considered here, $|\Theta_{p^*,p'}^{(2)}|$ is roughly proportional to $\tau$. Similar calculations show that $|\Theta_{p,p'}^{(2)}|$ is roughly proportional to $\tau$ when $p \neq p^*$, $p' = p^*$ and $p = p' \neq p^*$ as well. 

Second, consider $p=p'=p^*$. Then, from (\ref{eq:theta_ppnn_2_lim}), we take the limit $\omega_{p',n'}\tau \rightarrow 0$ again and obtain 
\begin{equation*}
   \Theta_{p^*,p^*,n,n'}^{(2)} \rightarrow - e^{-i \omega_{p',n'}\tau/4} A_n A^*_{n'} \frac{\tau^2}{4}.
\end{equation*}
Thus, $|\Theta_{p^*,p^*}^{(2)}|$ is roughly proportional to $\tau^2$.

Third, consider $p \neq p^*$ and $p' \neq p^*$ where $p \neq p'$. In such case, for all $n$ and $n'$ such that $|A_n|$ and $|A_{n'}|$ are both non-negligible, $\omega_{p,n}$, $\omega_{p',n'}$, and $\omega_{p,n} - \omega_{p',n'}$ in (\ref{eq:theta_ppnn_2}) are all nonzero. Thus, as $\tau$ is increased, we expect from (\ref{eq:theta_ppnn_2}) that $|\Theta_{p,p'}^{(2)}|$ oscillates with a roughly constant amplitude ($\propto \tau^0$). 

\begin{figure*}[ht!]
\includegraphics[width=\textwidth]{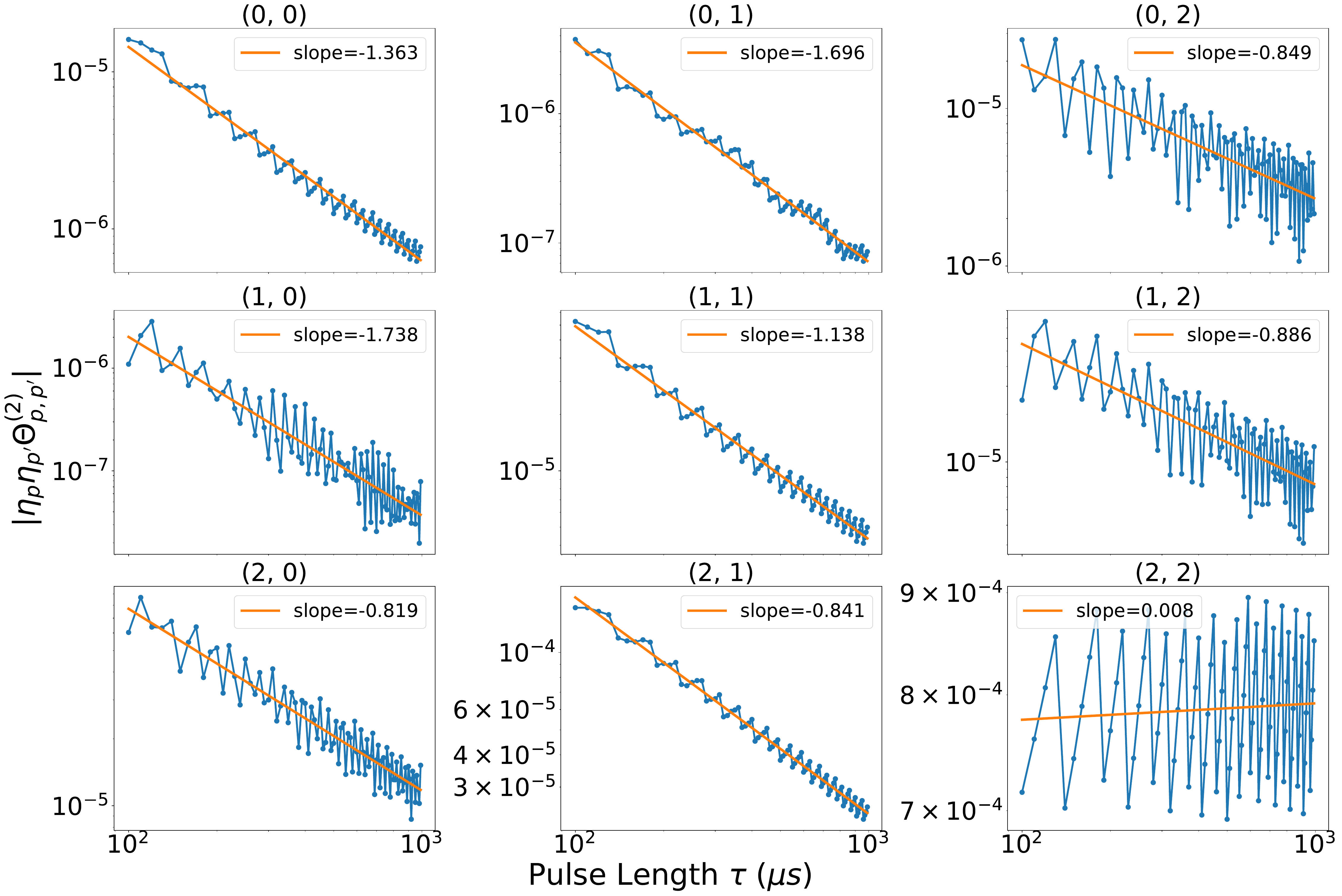}
\caption{Scaling trends of the second-order Magnus integrals $|\Theta_{p,p'}^{(2)}|$ (multiplied by $|\eta_p \eta_{p'}|$) with respect to the pulse length $\tau$, for various mode pairs $(p,p')$. Here, moment-0 shaped pulses with $\alpha=1$ are used on a 3-ion, 3-mode chain, where mode 2 is the target mode.}
\label{fig:sec_mag}
\end{figure*}

Figure~\ref{fig:sec_mag} plots the values of $|\eta_p \eta_{p'}\Theta_{p,p'}^{(2)}|$ of the moment-0 shaped pulses targeting mode $p^*=2$ for various pulse lengths. We emphasize that $\alpha$ is fixed to 1 in these plots. As $\alpha \approx \bar{A} \tau$ for moment-0 shaped pulses (see Appendix~\ref{sec:appendix_first_order_magnus}), the proportionality to $\tau^\beta$ when $\bar{A}$ is fixed shows up as the proportionality to $\tau^{\beta-2}$ in these plots, where the additional proportionality to $\tau^{-2}$ comes from $|\Theta_{p,p'}^{(2)}| \propto \bar{A}^2$. As expected from the above discussion, with fixed $\alpha$, $|\Theta_{p,p'}^{(2)}|$ is roughly proportional to $\tau^{\beta-2}$ where the closest integer to $\beta$ is 2 for $(p,p') = (2,2)$, 0 for $(p,p') = (0,1)$ and $(1,0)$, and 1 for all other $(p,p')$. Excluding $\Theta^{(2)}_{2,2}$ as it does not contribute to the CMC error, the largest second-order contribution to the CMC error comes from $\Theta^{(2)}_{2,1}$, whose magnitude is roughly proportional to $\tau^{\beta-2} = \tau^{-0.841}$. As explained above for the case where $p=p^*$ and $p' \neq p^*$, the value of $\beta$ is reasonably close to 1. 

Finally, we consider the scaling trend with respect to $\delta_p$. Similarly to $\Theta_p^{(1)}$, in the presence of detuning $\delta_p$, there are contributions to $\Theta_{p,p'}^{(2)}$ from all orders of $\delta_p$. For the pulse-shaping method in the main text, only  $\Theta_p^{(1)}$ is stabilized with respect to $\delta_p$ and not $\Theta_{p,p'}^{(2)}$. Thus, the leading-order contribution of $\delta_p$ to $|\Theta_{p,p'}^{(2)}|$ is first order ($\propto \delta_p$).

\subsection{Mangus terms and the qubit poulation}
\label{sec:appendix_population}

In this subsection, we explain how the leading-order Magnus terms affect the qubit population. Combined with the previous two subsections, this completes the discussion on how the qubit population and its error scale with respect to parameters such as $\alpha$, $\tau$, and $\delta_p$. 

First, we consider the first-order Magnus term in (\ref{eq:Omega_1_1}). To leading order in $\eta_p$, we may write $\exp(\hat{\Omega}_1) \approx \prod_p \exp(\hat{\Omega}_{1,p})$, where
\begin{equation}
    \hat{\Omega}_{1,p} = \eta_p \Theta_p^{(1)} \hat{\sigma}^+ \hat{a}_p^\dagger - h.c.
\end{equation}
Thus, the first-order contribution of mode $p$ to the qubit population is induced by the unitary $\exp(\hat{\Omega}_{1,p})$, followed by projection to the qubit subspace. In this paper, we set $|0,0\rangle$ as the initial state, where $|a,b\rangle$ is the composite state of the qubit state ($a=0,1$) and the motional Fock state ($b=0,1,2,...$). In such case, $\hat{\Omega}_{1,p}$ induces transition between only two states $|\tilde{0}\rangle := |0,0\rangle$ and $|\tilde{1}\rangle := |1,1\rangle$. Specifically, $\hat{\Omega}_{1,p} = \eta_p \Theta_p^{(1)} |\tilde{1} \rangle \langle \tilde{0}| - h.c. = i \eta_p |\Theta_p^{(1)}| (\sin \varphi \hat{\sigma}^{\tilde{x}} - \cos \varphi \hat{\sigma}^{\tilde{y}})$, where $\Theta_p^{(1)} = |\Theta_p^{(1)}| e^{i \varphi}$ and $\hat{\sigma}^{\tilde{x}}$ ($\hat{\sigma}^{\tilde{y}}$) is the Pauli-X (Y) operator of the qubit spanned by $|\tilde{0}\rangle$ and $|\tilde{1}\rangle$. Then, straightforward calculations show that
\begin{equation}
    \langle \tilde{1} | \exp(\hat{\Omega}_{1,p}) |\tilde{0}\rangle 
    = \langle \tilde{1} | \exp\left( i \eta_p |\Theta_p^{(1)}| (\sin \varphi \hat{\sigma}^{\tilde{x}} - \cos \varphi \hat{\sigma}^{\tilde{y}}) \right) |\tilde{0}\rangle 
    =  e^{i \varphi}
     \sin \left( \eta_p |\Theta_p^{(1)}| \right).
\end{equation}
In the perturbative regime $\eta_p |\Theta_p^{(1)}| \ll 1$ $\forall p$, when considering the contribution of mode $p$ to the qubit population, the qubit-state transition due to other modes can be ignored. Thus, we conclude that the magnitude of first-order contribution of mode $p$ to the qubit's bright-state ($|1\rangle$) population is approximately given by $\sin^2 ( |\eta_p \Theta_p^{(1)}| ) \approx |\eta_p \Theta_p^{(1)}|^2$. The largest contribution comes from the target mode $p^*$, which gives the approximate bright-state population
\begin{align}
    \mathcal{P} \approx \sin^2 (\eta_{p^*} |\Theta_{p^*}^{(1)}|).
    \label{eq:pop}
\end{align}

Next, we consider the second-order Magnus term in (\ref{eq:sec_order_term}).
Similarly to above, we define
\begin{equation}
    \hat{\Omega}_{2,p,p'} = \eta_p \eta_{p'} \Theta^{(2)}_{p,p'} (\hat{\sigma}^z \hat{a}_p^\dagger \hat{a}_{p'} + \hat{\sigma}^- \hat{\sigma}^+ \delta_{p,p'}) - h.c.,
\end{equation}
and as a concrete example, we analyze the contribution of 
\begin{equation}
    \hat{\Omega}_{2,p^*,p^*}
    = 2i \eta_{p^*}^2 \Im[\Theta^{(2)}_{p^*,p^*}] (\hat{\sigma}^z \hat{a}_{p^*}^\dagger \hat{a}_{p^*} + \hat{\sigma}^- \hat{\sigma}^+)
\end{equation}
to the qubit population, which adds perturbatively to the contribution of $\hat{\Omega}_{1,p^*}$. An important observation is that in the subspace spanned by $|\tilde{0}\rangle$ and $|\tilde{1}\rangle$, $\hat{\sigma}^z \hat{a}_{p^*}^\dagger \hat{a}_{p^*} + \hat{\sigma}^- \hat{\sigma}^+ = |\tilde{0}\rangle \langle \tilde{0}| - |\tilde{1}\rangle \langle \tilde{1}| = \hat{\sigma}^{\tilde{z}}$. Again by straightforward calculations
\begin{align}
    \langle \tilde{1} | \exp(\hat{\Omega}_{1,p^*} + \hat{\Omega}_{2,p^*,p^*}) |\tilde{0}\rangle
    = \langle \tilde{1} | \exp \left( i B (\sin \varphi \hat{\sigma}^{\tilde{x}} - \cos \varphi \hat{\sigma}^{\tilde{y}}) + i C \hat{\sigma}^{\tilde{z}} \right)|\tilde{0}\rangle
    = \frac{Be^{i \varphi}}{\sqrt{B^2+C^2}} \sin \sqrt{B^2 + C^2},
\end{align}
where $B = \eta_{p^*} |\Theta_{p^*}^{(1)}|$ and $C = 2 \eta_{p^*}^2 \Im[\Theta_{p^*,p^*}^{(2)}]$. Therefore, the contribution of $\hat{\Omega}_{1,p^*} + \hat{\Omega}_{2,p^*,p^*}$ to the qubit population is given by $\frac{B^2}{B^2+C^2} \sin^2 \sqrt{B^2+C^2}$, which is approximately $B^2 - C^2$ in the perturbative regime where the expansion is performed with respect to $C/B \ll 1$ first and then $B \ll 1$. Compared to the case where the qubit-population transition is induced only by $\hat{\Omega}_{1,p^*}$, the addition of $\hat{\Omega}_{2,p^*,p^*}$ reduces the bright-state population by $C^2 = (\Im[\Theta^{(2)}_{p^*,p^*}])^2$. Other second-order Magnus terms $\hat{\Omega}_{2,p,p'}$ [$(p,p') \neq (p^*,p^*)$] are also expected to induce changes in the bright-state population that scale as $|\Theta^{(2)}_{p,p'}|^2$, as $\hat{\Omega}_{2,p,p'}$ is proportional to either $\hat{\sigma}^z$ (when $p \neq p'$) or $\hat{\sigma}^{\tilde{z}}$ (when $p = p'$).

Combined with the analysis in Appendix~\ref{sec:appendix_first_order_magnus} and \ref{sec:appendix_second_order_magnus}, this completes the discussion on how the first- and second-order contributions to the qubit population scale with respect to parameters such as $\alpha$, $\tau$, and $\delta_p$. For example, the scaling behaviors of $\mathcal{E} = |P-\mathcal{P}|/\mathcal{P}$ with respect to $\alpha$ and $\tau$ in Fig.~\ref{fig:2} is now fully understood. The denominator $\mathcal{P}$ in the $|\eta_{p^*} \alpha| \ll 1$ regime is approximately given by (\ref{eq:pop}) as $\sin^2(\eta_{p^*} \alpha) \approx (\eta_{p^*} \alpha)^2$, which is proportional to $\alpha^2$ in Fig.~\ref{fig:2}(a) and fixed to a constant in Fig.~\ref{fig:2}(b). For the numerator $|P - \mathcal{P}|$, the leading contribution is from $|\Theta_{p'}^{(1)}|^2$ ($|\Theta_{p^*,p'}^{(2)}|^2$) where $p' \neq p^*$ for square (shaped) pulses, which is roughly proportional to $\bar{A}^2 \tau^0$ ($\bar{A}^4 \tau^2$). Thus, in Fig.~\ref{fig:2}(a), for fixed $\tau \approx \alpha / \bar{A}$, $\mathcal{E}$ scales as $\bar{A}^2 / \alpha^2 \propto \alpha^0$ ($\bar{A}^4/\alpha^2 \propto \alpha^2$) for square (shaped) pulses. Also, in Fig.~\ref{fig:2}(b), for fixed $\alpha \approx \bar{A}\tau$, $\mathcal{E}$ scales as $\bar{A}^2 \tau^0 \propto \tau^{-2}$ ($\bar{A}^4 \tau^2 \propto \tau^{-2}$) for square (shaped) pulses. Similar arguments can be made for the scaling with respect to $\delta_p$. These scaling behaviors are important to understand the regime of parameters in which shaped pulses may allow more accurate mode characterization than square pulses.

\subsection{Even- vs. odd-moment stabilization}
\label{app:even_odd_stabilization}
In Fig.~\ref{fig:3}(c) and (d), we observe that for integer $K$ and a wide range of pulse length $\tau$ and detuning $\delta$, moment-$2K$ pulses in general perform better than moment-$(2K+1)$ pulses. For example, while Fig.~\ref{fig:3}(b) clearly shows that moment-1 pulse suppresses $|\Theta_{p^*}^{(1)}(\delta) - \Theta_{p^*}^{(1)}(0)|$ compared to moment-0 pulse, Fig.~\ref{fig:3}(d) shows that moment-1 pulse has larger qubit-population error $\mathcal{E}(\delta)$ than moment-0 pulse for all plotted $\delta$. This appendix provides an explanation for this phenomenon. The key understanding is that while our pulse-shaping method for stabilization can suppress $|\Theta_p^{(1)}(\delta) - \Theta_p^{(1)}(0)|$, this is not equivalent to suppressing $| |\Theta_p^{(1)}(\delta)|^2 - |\Theta_p^{(1)}(0)|^2 |$, which is more relevant for the error in qubit population that is approximately given by $|\eta_{p^*} \Theta_{p^*}^{(1)}|^2$. 

For simplicity, we consider a resonant square pulse $g(t) = \bar{A} e^{-i\omega_{p^*}t + i\phi}$ as an example, and without losing generality we set $\phi=0$ as global phase of the pulse does not affect the qubit population. We evaluate $\Theta_{p^*}^{(1)}$ as this is the dominant contribution to the qubit population. From (\ref{eq:cmc}), we can easily find that $\partial^k \Theta_{p^*}^{(1)} / \partial \omega_{p^*}^k$ is real (imaginary) when $k$ is even (odd). In the presence of detuning $\omega_{p^*} \rightarrow \omega_{p^*} + \delta$, this leads to
\begin{equation}
    |\Theta_{p^*}^{(1)}(\delta)|^2 = 
    \left( \Theta_{p^*}^{(1)}(0) + \frac{\delta^2}{2!} \frac{\partial^2 \Theta_{p^*}^{(1)}}{\partial \omega_{p^*}^2} + \cdots \right)^2
    + \left( \delta \frac{\partial \Theta_{p^*}^{(1)}}{\partial \omega_{p^*}} + \frac{\delta^3}{3!} \frac{\partial^3 \Theta_{p^*}^{(1)}}{\partial \omega_{p^*}^3} + \cdots \right)^2,
\end{equation}
where all derivatives are evaluated at $\delta = 0$. Thus, the qubit-population error due to detuning $\delta$ is proportional to
\begin{equation}\label{eq:evenvsoddgeneral}
    |\Theta_{p^*}^{(1)}(\delta)|^2 - |\Theta_{p^*}^{(1)}(0)|^2
    = \sum_{k=1}^\infty \left[
    \frac{2}{(2k)!} \Theta_{p^*}^{(1)}(0) \frac{\partial^{2k} \Theta_{p^*}^{(1)}}{\partial \omega_{p^*}^{2k}}
    + \sum_{l=1}^{k-1} \frac{2}{l!(2k-l)!} \frac{\partial^{l} \Theta_{p^*}^{(1)}}{\partial \omega_{p^*}^{l}}\frac{\partial^{2k-l} \Theta_{p^*}^{(1)}}{\partial \omega_{p^*}^{2k-l}}
    + \left( \frac{1}{k!} \frac{\partial^k \Theta_{p^*}^{(1)}}{\partial \omega_{p^*}^k} \right)^2
    \right] \delta^{2k},
\end{equation}
which leads to
\begin{equation}\label{eq:evenvsodd}
    |\Theta_{p^*}^{(1)}(\delta)|^2 - |\Theta_{p^*}^{(1)}(0)|^2
    =  \left[ \Theta_{p^*}^{(1)}(0) \frac{\partial^{2} \Theta_{p^*}^{(1)}}{\partial \omega_{p^*}^{2}} + \left( \frac{\partial \Theta_{p^*}^{(1)}}{\partial \omega_{p^*}} \right)^2 \right] \delta^2 + \mathcal{O}(\delta^4)
\end{equation}

While (\ref{eq:evenvsodd}) is derived for square pulse, we argue that it is approximately valid for shaped pulses that have similar spectrum to square pulse, such as moment-0 and 1 pulses in Fig.~\ref{fig:pulse_prof}. In such case, (\ref{eq:evenvsodd}) explains why moment-1 pulse does not lead to smaller $\mathcal{E}(\delta)$ than moment-0 pulse. Moment-1 pulse only imposes $\partial \Theta_{p^*}^{(1)} / \partial \omega_{p^*} = 0$, which removes only the second term of the right-hand-side of (\ref{eq:evenvsodd}); the first term survives. Thus, the leading-order contribution of $\delta$ to $\mathcal{E}(\delta)$ is $\mathcal{O}(\delta^2)$ for both moment-0 and moment-1 pulses. At least second-order stabilization, which imposes $\partial^k \Theta_{p^*}^{(1)} / \partial \omega_{p^*}^k = 0$ for $k=1$ and $2$, is required to completely remove the $\mathcal{O}(\delta^2)$ contribution to $\mathcal{E}(\delta)$. As moment-1 pulses are more susceptible to the CMC than moment-0 pulses due to the larger average Rabi frequency $\bar{A}$ [see Fig.~\ref{fig:3}(a)], we expect moment-1 pulses to have larger qubit-population error than moment-0 pulses. 

The higher-order terms in (\ref{eq:evenvsoddgeneral}) suggest a more general claim that moment-$(2K+1)$ pulse does not outperform moment-$2K$ pulse in terms of smaller $\mathcal{E}(\delta)$. Indeed, Fig.~\ref{fig:4} shows that moment-3 pulses do not outperform moment-2 pulses for a significant range of $|\delta|$, especially for shorter pulse lengths. However, as shown in Fig.~\ref{fig:pulse_prof}, pulses with moment 2 or higher tend to have a spectrum that is substantially different from the resonant square pulse. Thus, for higher-moment pulses, $\mathcal{O}(\delta^{2K+1})$ terms may arise in the right-hand-side of (\ref{eq:evenvsoddgeneral}), which are completely removed by moment-$(2K+1)$ pulses but not by moment-$2K$ pulses. This is evidenced in the large-$|\delta|$ region of Fig.~\ref{fig:4}  where moment-3 pulses outperform moment-2 pulses. Whether moment-2 or moment-3 pulse leads to an overall better performance depends on the expected probability distribution of $\delta$ and the feasible range of pulse length $\tau$.

\section{Nulling Second-order Magnus Integrals}
\label{app:nulling_second_magnus}
To further minimize the CMC error, we describe our attempt at nulling the second-order Magnus integrals, thereby silencing the second-order Magnus terms in (\ref{eq:sec_order_term}). We adopt a similar approach to the pulse-shaping scheme in the main text that nulls unwanted first-order Magnus integrals. The scheme proposed here can be used in conjunction to the pulse-shaping scheme in the main text.

Specifically, we write an $N_{\rm{basis}}\times N_{\rm{basis}}$ matrix $\mathbf{Q}^{(p,p')}$ for each mode pair $(p,p')\neq (p^*,p^*)$ whose element is given by
\begin{equation*}
    Q_{n,n'}^{(p,p')} = \Theta^{(2)}_{p,p',n,n'}.
\end{equation*}
We then find the singular value decomposition of $|\mathbf{Q}^{(p,p')}|^2$ and select the set of singular vectors $S_{(p,p')}$ with singular values below a sufficiently small threshold of, for example, $10^{-10}$. The second-order Magnus integral $\Theta^{(2)}_{p,p'}$ is nulled by the pulse represented by a vector in $S_{(p,p')}$. We iterate this process over all mode-pairs except $(p,p')=(p^*,p^*)$ and find the overlap $\boldsymbol{\mathcal{S}}^{(2)}$ between the vector spaces $\mathbf{S}_{(p,p')}$ spanned by the vectors in $S_{(p,p')}$, i.e., $\boldsymbol{\mathcal{S}}^{(2)}:=\bigcap_{p,p',(p,p')\neq (p^*,p^*)} \mathbf{S}_{(p,p')}$. Finally, we find the overlap between $\boldsymbol{\mathcal{S}}^{(1)}$ nulling $\Theta^{(1)}_{p}$ as discussed in Sec.~\ref{sec:method} and $\boldsymbol{\mathcal{S}}^{(2)}$ nulling $\Theta^{(2)}_{p,p'}$, and proceed with the signal strength maximization over the set of vectors spanning the final overlapping space $\boldsymbol{\mathcal{S}}^{(1)}\cap \boldsymbol{\mathcal{S}}^{(2)}$. 

To find the overlap between two vector spaces, for example, $\boldsymbol{\mathcal{S}}^{(1)}$ and $\boldsymbol{\mathcal{S}}^{(2)}$, with spanning vectors $\mathcal{S}^{(1)}$ and $\mathcal{S}^{(2)}$, respectively, we find the nullspace of the enlarged matrix $U = \left(\hat{\mathcal{S}}^{(1)}|-\hat{\mathcal{S}}^{(2)}\right)$, where $\hat{\mathcal{S}}^{(1)}$ ($\hat{\mathcal{S}}^{(2)}$) is the matrix whose columns are the vectors in $\mathcal{S}^{(1)}$ ($\mathcal{S}^{(2)}$). Let's denote the vectors spanning the nullsapce of $U$ as $n_i^T=(n_i^{(1)} |\: n_i^{(2)})^T$. Then, the intersection space $\boldsymbol{\mathcal{S}}^{(1)}\cap \boldsymbol{\mathcal{S}}^{(2)}$ is spanned by the vectors $\hat{\mathcal{S}}^{(1)}n_i^{(1)}=\hat{\mathcal{S}}^{(2)}n_i^{(2)}$.

In practice, we found that the matrix $\mathbf{Q}^{(p,p')}$ is positive-definite and does not have sufficiently small singular values when $p = p'$. Thus, our method above is applicable only when mode pairs ($p,p'$) such that $p \neq p'$ are considered. In Fig.~\ref{fig:sec_mag}, we show that for the case of a 3-ion chain and target mode $p^*=2$, the largest second-order contribution to the CMC error comes from $\Theta^{(2)}_{2,1}$. Thus, one might hope that additionally nulling the second-order Magnus integrals $\Theta^{(2)}_{p,p'}$ for $p \neq p'$ is sufficient for reducing the population error, compared to the pulse-shaping scheme in the main text that only nulls the first-order Magnus integrals. Unfortunately, it turns out that due to the increased power requirement for these additional nulling constraints, nulling $\Theta^{(2)}_{p,p'}$ for $p \neq p'$ results in increased magnitude of $\Theta^{(2)}_{p,p'}$ for $p = p'$, and thus no reduction in the qubit-population error. We expect similar results for all pulse lengths, as $|\Theta^{(2)}_{p^*,p'}|$ and $|\Theta^{(2)}_{p',p'}|$ ($p' \neq p^*$) exhibit similar scaling behavior with respect to $\tau$, as discussed in Appendix~\ref{sec:appendix_second_order_magnus}. 

\section{Comparison of Error Landscapes}
\label{app:ultimate_comparison}

\begin{figure*}[htb!]
\includegraphics[width=\textwidth]{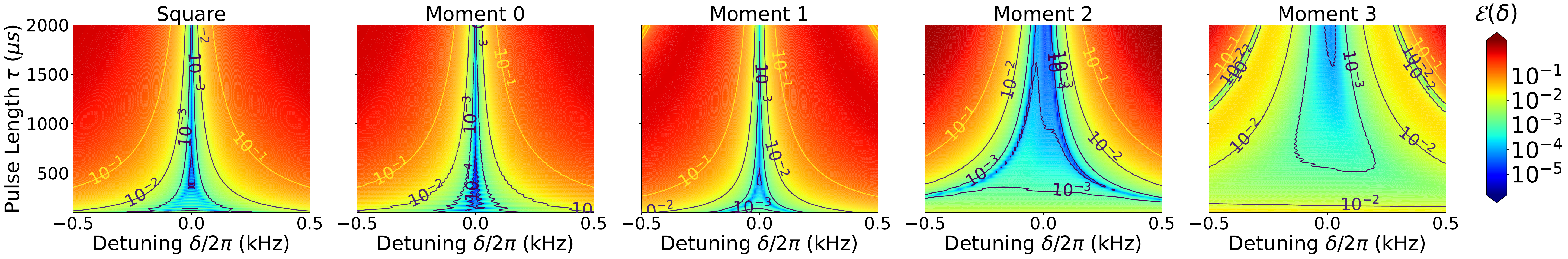}
\caption{Error landscapes of shaped pulses of $\alpha=1$ with cross-mode coupling and detuning error. The fractional error $\mathcal{E}(\delta) = |P(\delta)-\mathcal{P}(\delta=0)|/\mathcal{P}(\delta=0)$ is plotted on a logged color scale shared between all panels. Contour lines are plotted for fractional error of $10^{-4}$ to $10^{-1}$. 
}
\label{fig:error_landscape}
\end{figure*}

In Fig.~\ref{fig:4}, we compared the performance of square and shaped pulses with different moments of stabilization in various regimes of pulse length $\tau$ and mode-frequency detuning $\delta$. Shaped pulses of different moments achieve the smallest qubit-population error $\mathcal{E}(\delta)$ compared to other pulses in different parameter regimes. Figure~\ref{fig:error_landscape} plots the values of $\mathcal{E}(\delta)$, providing a full comparison between square and shaped pulses of various moments of stabilization. 

We first consider the case where $|\delta|$ is very small, such that the CMC is the dominant source of error $\mathcal{E}(\delta)$. In such case, as $\alpha$ is fixed and $\tau$ increases, the average Rabi frequency $\bar{A}$ decreases, which results in smaller effects of CMC and thus smaller $\mathcal{E}(\delta)$. This trend is observed in the narrow area at the horizontal center of each panel in Fig.~\ref{fig:error_landscape} as well as in Fig.~\ref{fig:2}(b). 

Next, we consider the case where $|\delta|$ is relatively large such that the detuning error dominates the CMC error. Here, an opposite trend emerges; as $\alpha$ is fixed and $\tau$ increases, $\mathcal{E}(\delta)$ increases. This is because as $\tau$ increases, a larger phase accumulates due to the detuning over the pulse duration. This trend is observed in the left and right ends of each panel in Fig.~\ref{fig:error_landscape} where $|\delta|$ is relatively large, as well as in Fig.~\ref{fig:3}(c). 

In the intermediate regime of $|\delta|$, effects of the CMC and detuning errors compete with each other. Given an expected range of detuning, or more specifically, the expected probability distribution of $\Pr(\delta)$, one can estimate the expected qubit-population error $\langle \mathcal{E} \rangle$ at each pulse length $\tau$ by integrating $\Pr(\delta)\mathcal{E}(\delta)$ over $\delta$. Then, the smallest $\langle \mathcal{E} \rangle$ will be achieved at an optimal $\tau$, where a transition from CMC-dominant regime ($\langle \mathcal{E} \rangle$ decreases as $\tau$ increases) to detuning-dominant regime ($\langle \mathcal{E} \rangle$ increases as $\tau$ increases) occurs. This transition is most clearly shown in Fig.~\ref{fig:error_landscape} for moment-2 and 3 shaped pulses and the plotted range of $\delta$, and also shown for lower-moment and square pulses when a narrower range of $\delta$ (e.g., $|\delta| \lesssim 2\pi \times 0.05 $ kHz) is considered. The optimal pulse length $\tau$ that achieves the smallest $\langle \mathcal{E} \rangle$ depends on the distribution $\Pr(\delta)$. This observation provides a general guideline for choosing the type of pulse and its length that enables as accurate mode characterization as possible. 

\section{Applicability of Our Proposed Method for Longer Ion Chains}
\label{app:applicability}

For longer ion chains, silencing the CMC effect is expected to be more challenging. As the number of motional modes increases, more nulling constraints need to be satisfied when finding the pulse solution. Also, assuming equidistantly spaced ions, the spacings between neighboring mode frequencies are typically smaller for longer ion chains, as shown in Table~\ref{tab:modespacings}. With the more tightly-spaced motional mode frequencies, it is reasonable to expect that nulling $\Theta_{p}^{(1)}$ for all $p \neq p^*$ requires more pulse resources such as power and time duration. This brings two main concerns for the scalability of our pulse shaping scheme, namely the runtime and the power requirement. In this appendix, we address both of these concerns by solving for moment-0 shaped pulses for ion chains of length up to 7, demonstrating the scalability of our proposed pulse-shaping scheme. 

\begin{table*}[ht]
\renewcommand*{\arraystretch}{1.5}
\begin{tabular}{ c | c | c | c | c | c | c }
\hline
$N$ & $\Delta_{0,1}$ & $\Delta_{1,2}$ & $\Delta_{2,3}$ & $\Delta_{3,4}$ & $\Delta_{4,5}$ & $\Delta_{5,6}$\\
\hline
3 & 96.8 & 68.0 &  & & & \\
4 & 80.3 & 63.1 & 43.9 & & & \\
5 & 62.9 & 53.2 & 42.8 & 29.2 & & \\
6 & 53.9 & 48.2 & 41.6 & 34.4 & 21.8 & \\
7 & 43.6 & 41.5 & 38.4 & 34.1 & 29.2 & 15.9 \\
\hline
\end{tabular}
\caption{Spacings between neighboring mode frequencies $(\omega_{p+1}-\omega_{p})/2\pi$ are given in units of kHz, for various numbers of ions (modes) $N=N'$. All values of mode parameters are obtained by numerically
solving the normal modes of equidistantly spaced ions
trapped by a modelled potential of an HOA2.0 trap \cite{HOA}.} \label{tab:modespacings}
\end{table*}

\begin{figure*}[htb!]
\includegraphics[width=0.8\textwidth]{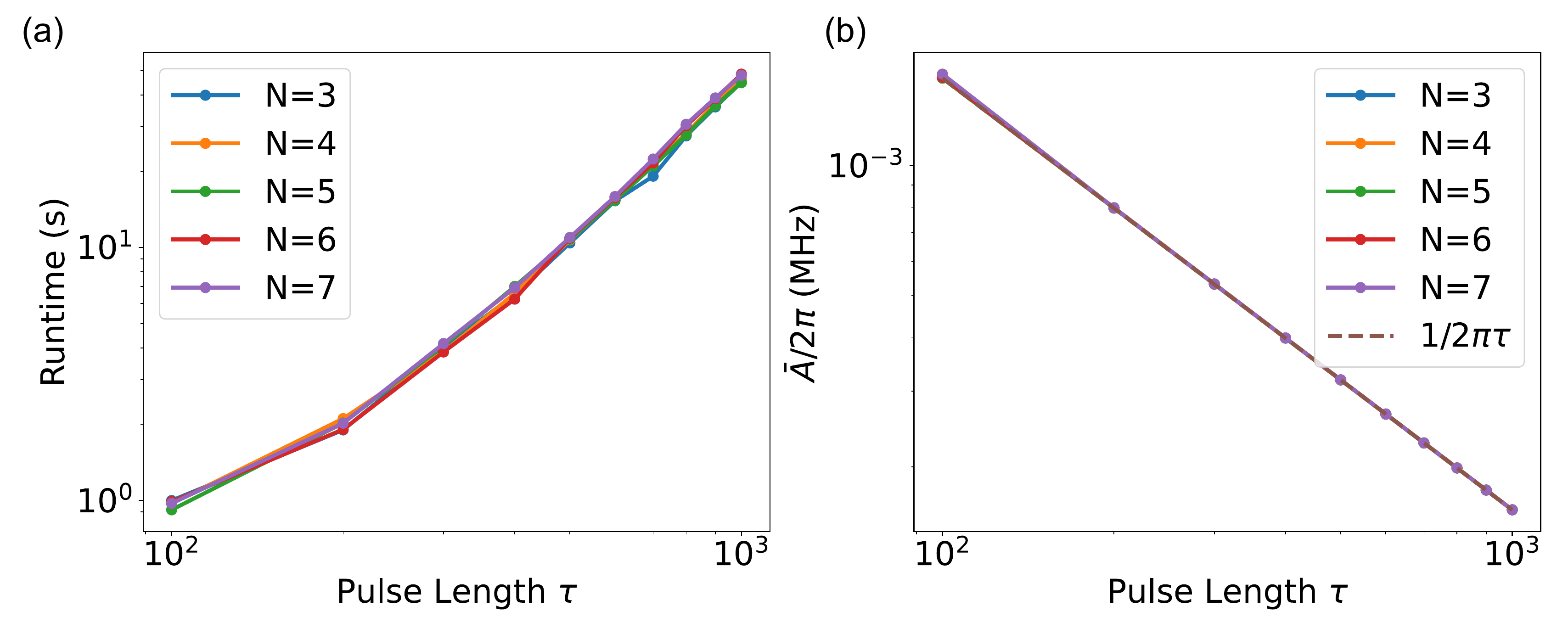}
    \caption{CPU runtime for pulse solver (a) and average Rabi frequency $\bar{A}$ (b) for shaped pulses of various lengths $\tau$ on ion chains with various numbers of ions/modes $N=N'$. We solve for moment-0 pulses with $\alpha=1$. Here we use the values of $\omega_p$'s and $\eta_{j,p}$'s from Appendix E of Ref.~\cite{Mingyu2022}. }
\label{fig:scalability}
\end{figure*}

First, we show how the runtime for finding the pulse solution scales with the number of ions (modes) $N=N'$. Specifically, we report the CPU runtime of our pulse solver on an Apple M1 Max chip. Although the number of nulling constraints increases linearly with $N$, we witness no increase in the runtime as $N$ is increased from 3 to 7, as shown in the left panel of Fig.~\ref{fig:scalability}. Instead, the runtime significantly increases with the pulse length $\tau$, which is proportional to the number of bases $N_{\rm{basis}}$ used. This is because $N_{\rm{basis}}$ determines the size of the matrix of which the nullspace vectors are calculated, which takes the longest runtime in our pulse-shaping scheme. We note that finding pulse solutions with length as long as $\tau = 2000$ $\mu$s only takes runtime less than a minute. As the runtime does not increase with $N$ (at least up to 7), we conclude that the runtime is unlikely to be a bottleneck for the scalability of our pulse-shaping scheme. 

Next, we discuss how the average Rabi frequency $\bar{A}$ of the shaped pulse scales with $N$. Somewhat surprisingly, the average Rabi frequency of moment-0 pulse does not change significantly as the number of modes is increased from 3 to 7, as shown in the right panel of Fig.~\ref{fig:scalability}. At least for the simulated numbers of modes and mode-frequency values, adding the nulling constraint $\Theta_{p}^{(1)}=0$ for $p \neq p^*$ does not incur additional power requirement, compared to square pulse with $\bar{A} = \alpha/\tau$. This is consistent with our previous observation in Fig.~\ref{fig:3}(a) that $\bar{A}$ is essentially the same for square and moment-0 pulses and increases only with larger moment of stabilization. The fact that suppressing the CMC effect is ``free'' in terms of power requirement makes our pulse-shaping scheme even more promising for application to longer ion chains. 

\end{document}